\documentclass[]{aa}
\usepackage{natbib,twoopt}
\usepackage{amsmath}
\usepackage{wasysym}
\usepackage{multirow}
\usepackage{longtable}
\usepackage{graphicx}
\usepackage{mathtools}
\usepackage{blindtext}
\usepackage[varg]{txfonts}

\bibpunct{(}{)}{;}{a}{}{,} 

\newcommand{\Ha}{H$_\alpha$}
\newcommand{\Haexcess}{$\Delta H\alpha$}
\newcommand{\Ham}{$\langle H_\alpha \rangle$}
\newcommand{\sig}{$\sigma$}
\newcommand{\RHK}{$\log R'_{HK}$}

\title{\Ha-activity and ages for stars in the SARG survey\thanks{Based on observations made with the Italian Telescopio Nazionale Galileo (TNG) operated 
on the island of La Palma by the Fundacion Galileo Galilei of the INAF (Istituto Nazionale di 
Astrofisica) at the Spanish Observatorio del Roque de los Muchachos of the Instituto de 
Astrofisica de Canarias}$^,$\thanks{Table 5 is only available in electronic form
at the CDS via anonymous ftp to cdsarc.u-strasbg.fr (130.79.128.5)
or via http://cdsweb.u-strasbg.fr/cgi-bin/qcat?J/A+A/} }
\author{E. Sissa\inst{1,2} 
	\and R. Gratton \inst{1}
	\and S. Desidera\inst{1}
	\and A. F. Martinez Fiorenzano\inst{3} 
	\and A. Bonfanti \inst{2}
	\and E. Carolo \inst{1} 
	\and D. Vassallo \inst{1,2}
	\and R.U. Claudi \inst{1}
	\and M. Endl \inst{4}
	\and R. Cosentino \inst{3,5 }}
\authorrunning{E. Sissa et al.}
\offprints{E. Sissa,  \\
   \email{elena.sissa@oapd.inaf.it} }
\institute{INAF - Osservatorio Astronomico di Padova,  
             Vicolo dell'Osservatorio 5, I-35122, Padova, Italy
             \and 
             Dipartimento di Fisica e Astronomia - Universita' di Padova, Vicolo
             dell'Osservatorio 3, I-35122, Padova, Italy 
             \and
             Fundaci\'on Galileo Galilei - INAF,
             Rambla Jos\'e Ana Fernandez P\'erez, 7
             38712 Bre\~na Baja, TF, Spain
             \and
             McDonald Observatory, The University of Texas at Austin, Austin, 
             TX 78712, USA
             \and
             INAF - Osservatorio Astrofisico di Catania, Via S.Sofia 78, Catania, Italy}

\date{Received  /
Accepted }

\abstract{Stellar activity influences radial velocity (RV) measurements and can also  mimic the presence of orbiting planets. As part of the search for planets around the components of wide binaries performed with the SARG High Resolution Spectrograph  at the
TNG, it was discovered that \object{HD 200466A} shows strong variation in RV that is well correlated with the activity index based on \Ha.  
We used SARG to study the \Ha\ line variations in each component of the binaries and a few bright stars to test the capability of the \Ha\ index  of revealing the rotation period or activity cycle. We also analysed the relations between the average activity level and other physical properties of the stars. We finally tried to reveal signals in the RVs that are due to the activity.
At least in some cases the variation in the observed RVs is due to the stellar activity. We confirm that \Ha\ can be used as
an activity indicator for solar-type stars and as an age indicator for stars younger than 1.5 Gyr.}

\begin{document}
\maketitle

\section{Introduction}
Studying the variation in the radial velocity (RV) induced by the chromospheric activity is important to distinguish it from the Keplerian motion of the star that may be caused by a planet \citep[see e.g.][]{Queloz01, Dumusque11, 2014Robertson}. On long timescales the active regions can modify measured RVs by introducing a signal related to the stellar activity cycle, while on short timescales the rotational period can become evident.

The most  widely used activity indicators are based on the Ca II H\&K lines \citep{2010Isaacson, Lovis11, Gomes11}, which have been shown to correlate with the radial velocity jitter. 
Other lines were investigated and it was found that the \Ha\ line can be a good alternative 
\citep{ 1990Robinson, 1990Stassmeier,2010Santos, Gomes11}.
However the correlation of \Ha\ with Ca II H\&K indices is high for the most active stars but decreases at a lower activity level, and sometimes becomes an anti-correlation  \citep{Gomes11}.
Similar results were also found by \cite{2007Cincunegui}, who added, using simultaneous observations of stars with spectral type later than F, that the correlation is lost when studying individual spectra of single stars and there is no dependence on activity. 
The correlation between the averaged fluxes for the Ca II and \Ha\ lines can be clarified by considering the dependence of the two indexes on the stellar colour or the spectral type, while the absence of a general relation between the simultaneous Ca II and \Ha\ index can be 
due to difference in the formation region of the two lines  \citep{2007Cincunegui, Gomes14}. Studying the solar spectrum as a prototype and extrapolating the results to other stars, \cite{2009Meunier} discovered that plages and 
filaments in the chromosphere contribute differently to Ca II and \Ha\ lines: while plages contribute to the emission of all these lines, the absorption due to filaments is remarkable only for \Ha. Therefore the saturation of the plage filling factor seems to enhance the correlation between the two indexes in case of high stellar activity and low filament contribution. On the other hand, the anti-correlation between the emission in Ca II and \Ha\ for low active stars seems to depend only on a strong filament contrast if the filaments are well correlated with plages \citep[see also][]{Gomes14}.

A search for planets around the components of wide binaries was performed using SARG (Spettrografo Alta Risoluzione Galileo) at the Telescopio Nazionale Galileo (TNG) in the past years. Two planetary companions were detected  around \object{HD 132563B} and \object{HD 106515A} \citep{Desidera11, Desidera12}. 
\citet{Carolo14} found strong variations in the RVs of \object{HD 200466A} that could not be explained by  a stable planetary system, but which were well correlated to a \Ha\ based activity indicator, showing that they are due to an $\sim$1100-day activity cycle.
Stimulated by this finding,  we started a systematic analysis of \Ha\  in the binaries of the SARG sample to identify activity-induced RV variations and distinguish them  from  planetary signatures.
 We report here on the main results of the activity study made within this survey.
 We also include the measurements for additional stars observed by our group for other programs carried out with SARG.\\

\section{Observation and data reduction}
\label{sec:data}
SARG is the High Resolution Spectrograph at TNG, now decommissioned, which worked for about 12 years beginning in 2000 \citep{Gratton01}. 
The SARG survey was the first planet research program entirely dedicated to binary systems and aimed to determine the frequency of giant planets up to a few AU separation from their star in nearly equal-mass visual binaries using high-precision radial velocities.
 The sample of the survey included 47 pairs of stars from the Hipparcos Multiple Star Catalog \citep{1997Perryman}, considering binaries in the magnitude range 7.0 $<$ V $<$ 10.0, with magnitude differences between the components  $\Delta$V $<$  1.0, projected separations larger than 2" (to avoid strong contamination of the spectra), 
 parallaxes larger than 10 mas, and errors smaller than 5 mas, with B - V $>$ 0.45 mag and spectral types later than F7. For more details on the sample see \cite{Desidera2007}.
 The stars are typically at distance $<$ 50 pc from the Sun.  
 
 Between September 2000 and  April 2012 we collected up to 81 spectra per star with a typical exposure time of 900 s for a total amount of more than 6000 science images.
 
 In this work we also include six bright stars that were observed with SARG looking to search for hot-Neptunes orbiting planets \citep{GrattonHotNeptunes}. For these stars the integration time was set at 600 s except for \object{61Cyg B} and \object{40 Eri}, for which it was shorter to avoid saturation of the images because of their higher luminosity.
 \object{$\tau$ Cet}, \object{51 Peg} and \object{$\rho$ CrB} were used as RVs reference stars during the survey,  and thei signal-to-noise ratio (S/N) is typically greater then 270. In addition, \object{HD 166435} was observed as benchmark active star \citep{2005Martinez}. We decided to include this star in our sample as well.
 
Our data set therefore consists made of two sub samples: the binary sample and the bright stars sample.
 The first is unbiased with respect to activity (except for \object{HD 114723}, which was excluded because of its high rotation), while the latter has a bias toward low-activity stars except for \object{HD 166435}.
For all the observations we used  the SARG Yellow Grism (spectral range 4600-7900 \AA) and the 0.27 arcsec slit to obtain a resolution $R=144000$ with a $2\times1$ pixel binning. 
The observed spectral range was covered by two chips. 
The blue chip included the spectral range used for the RV determination: the accuracy was given by a iodine cell superimposing a forest of absorption lines used as reference for the AUSTRAL code \citep{Endl2000}, as shown in  \citet{Desidera11}. 
The red chip data are affected by fringing effects at wavelengths longer than $\sim7000$ \AA; these were not used in our analysis. The depth of the iodine lines decreases toward longer wavelengths, and the lines are negligible at the wavelength of \Ha.

Data reduction was performed with standard IRAF\footnote{\cite{IRAF}} procedures.

\section{Stellar parameters}
\label{sec:parameters}
\begin{longtab}

\begin{longtable}{l*{5}{c}rc}
\caption{\label{additionalparam}\textbf{Stellar Parameters:} for each star we indicate the apparent V magnitude, the B-V color index, the method used to calculate \RHK and its value, the projected velocity $v\sin i$ and temperature. \RHK values were derived from direct measurement (D) of other authors, from the X-ray luminosity (X), or can be only an upper limits (U), still from X-ray luminosity data.}  \\
\hline\hline
Star & V & B-V  &  \RHK & method   & $v\sin i$  & $T_{eff}$ \\
  	&	&		&			&			&	[km/s]	 & [K] \\	
\hline
\endfirsthead
\caption{Continued.}\\
\hline\hline
 Star  & V & B-V &  $\log RHK$ & method   & $v\sin i$  & $T_{eff}$ \\
   	&	&		&			&			&	[km/s]	& [K] \\
 \hline
\endhead
\hline 
\endfoot

%

   BD 182366A &  9.370 &     0.83 &    -4.56 &    U &      2.0 &   5308 \\ 
   BD 182366B &  9.427 &     0.93 &    -4.57 &    U &      2.4 &   5290 \\ 
   BD 222706A &  9.594 &     0.62 &    -4.97 &    D &      2.3 &   5943 \\ 
   BD 222706B &  9.828 &     0.69 &    -4.51 &    U &      2.3 &   5674 \\ 
   BD 231978A &  9.395 &     0.83 &    -4.46 &    D &      3.5 &   4886 \\ 
   BD 231978B &  9.530 &     0.75 &    -4.44 &    U &      3.2 &   4911 \\ 
   HD 105421A &  7.827 &     0.51 &    -4.70 &    U &      4.6 &   6324 \\ 
   HD 105421B &  8.358 &     0.57 &    -4.65 &    U &      0.9 &   6102 \\ 
   HD 106515A &  7.960 &     0.79 &    -5.04 &    D &      1.7 &   5314 \\ 
   HD 106515B &  8.234 &     0.83 &    -5.07 &    D &      1.8 &   5157 \\ 
   HD 108421A &  8.870 &     0.90 &    -4.57 &    X &      2.6 &   4700 \\ 
   HD 108421B &  9.274 &     0.88 &    -4.53 &    X &      3.2 &   4779 \\ 
    HD 108574 &  7.418 &     0.56 &    -4.49 &    D &      4.9 &   6205 \\ 
    HD 108575 &  7.972 &     0.67 &    -4.43 &    X &      5.1 &   5895 \\ 
   HD 109628A &  9.073 &     0.57 &    -4.51 &    U &      3.0 &   6109 \\ 
   HD 109628B &  9.087 &     0.55 &    -4.50 &    U &      3.2 &   6127 \\ 
   HD 117963A &  8.639 &     0.55 &    -4.65 &    U &      6.2 &   6180 \\ 
   HD 117963B &  8.924 &     0.49 &    -4.61 &    U &      3.3 &   6097 \\ 
   HD 118328A &  9.147 &     0.62 &    -4.61 &    U &      0.7 &   5943 \\ 
   HD 118328B &  9.426 &     0.69 &    -4.59 &    U &      0.7 &   5887 \\ 
   HD 121298A &  8.604 &     0.50 &    -4.91 &    D &      1.9 &   6353 \\ 
   HD 121298B &  8.937 &     0.52 &    -4.87 &    D &      1.3 &   6266 \\ 
   HD 123963A &  8.758 &     0.62 &    -4.63 &    U &      1.6 &   5873 \\ 
   HD 123963B &  9.511 &     0.60 &    -4.55 &    U &      1.4 &   5438 \\ 
   HD 124054A &  8.399 &     0.58 &    -4.97 &    D &      2.7 &   6081 \\ 
   HD 124054B &  8.785 &     0.64 &    -5.02 &    D &      2.5 &   5896 \\ 
   HD 126246A &  7.466 &     0.54 &    -4.40 &    D &      7.9 &   6223 \\ 
   HD 126246B &  7.697 &     0.60 &    -4.51 &    D &      3.9 &   6074 \\ 
   HD 128041A &  8.059 &     0.71 &    -4.53 &    U &      3.4 &   5663 \\ 
   HD 128041B &  8.827 &     0.78 &    -4.51 &    U &      3.2 &   5192 \\ 
   HD 132563A &  8.948 &     0.54 &    -4.62 &    U &      3.9 &   6168 \\ 
   HD 132563B &  9.402 &     0.57 &    -4.62 &    U &      3.4 &   5985 \\ 
   HD 132844A &  9.022 &     0.55 &    -4.66 &    U &      3.4 &   5878 \\ 
   HD 132844B &  9.114 &     0.63 &    -4.61 &    U &      2.4 &   5809 \\ 
    HD 13357A &  8.180 &     0.67 &    -4.70 &    D &      1.7 &   5615 \\ 
    HD 13357B &  8.647 &     0.73 &    -4.61 &    D &      1.8 &   5341 \\ 
   HD 135101A &  6.656 &     0.69 &    -4.99 &    D &      2.3 &   5631 \\ 
   HD 135101B &  7.500 &     0.75 &    -5.07 &    D &      1.1 &   5491 \\ 
   HD 139569A &  8.482 &     0.54 &    -4.55 &    U &      8.5 &   6223 \\ 
   HD 139569B &  8.783 &     0.57 &    -4.52 &    U &      5.6 &   5922 \\ 
   HD 143144A &  8.856 &     0.62 &    -4.61 &    U &      1.9 &   5943 \\ 
   HD 143144B &  9.025 &     0.61 &    -4.59 &    U &      1.2 &   5894 \\ 
   HD 146413A &  9.260 &     0.88 &    -4.68 &    D &      2.1 &   4779 \\ 
   HD 146413B &  9.492 &     0.87 &    -4.60 &    X &      2.0 &   4818 \\ 
    HD 17159A &  8.775 &     0.54 &    -4.64 &    U &      3.4 &   6155 \\ 
    HD 17159B &  8.923 &     0.53 &    -4.62 &    U &      2.9 &   6051 \\ 
   HD 186858A &  8.368 &     0.96 &    -4.73 &    D &      3.6 &   4910 \\ 
   HD 186858B &  8.578 &     0.93 &    -4.62 &    X &      3.0 &   4885 \\ 
   HD 190042A &  8.755 &     0.73 &    -4.71 &    U &      3.5 &   5474 \\ 
   HD 190042B &  8.778 &     0.80 &    -4.72 &    U &      3.8 &   5406 \\ 
    HD 19440A &  7.874 &     0.47 &    -4.73 &    U &      4.5 &   6308 \\ 
    HD 19440B &  8.574 &     0.53 &    -4.66 &    U &      2.9 &   6108 \\ 
   HD 200466A &  8.399 &     0.74 &    -4.77 &    D &      2.0 &   5633 \\ 
   HD 200466B &  8.528 &     0.76 &    -4.69 &    D &      2.1 &   5583 \\ 
   HD 201936A &  8.648 &     0.48 &    -4.55 &    X &      8.8 &   6441 \\ 
   HD 201936B &  8.851 &     0.50 &    -4.53 &    X &     15.5 &   6452 \\ 
   HD 209965A &  7.980 &     0.55 &    -4.96 &    D &      4.2 &   6180 \\ 
   HD 209965B &  8.414 &     0.57 &    -4.59 &    U &      2.1 &   6115 \\ 
   HD 213013A &  8.982 &     0.81 &    -4.59 &    U &      1.7 &   5402 \\ 
   HD 213013B &  9.612 &     0.93 &    -4.53 &    U &      2.3 &   4990 \\ 
   HD 215812A &  7.275 &     0.64 &    -4.66 &    U &      1.3 &   5688 \\ 
   HD 215812B &  7.576 &     0.71 &    -4.64 &    U &      1.5 &   5586 \\ 
   HD 216122A &  8.062 &     0.58 &    -4.73 &    U &      6.5 &   6067 \\ 
   HD 216122B &  8.186 &     0.58 &    -4.71 &    U &      4.6 &   6066 \\ 
   HD 219542A &  8.174 &     0.64 &    -5.07 &    D &      2.1 &   5849 \\ 
   HD 219542B &  8.547 &     0.72 &    -4.81 &    D &      1.9 &   5691 \\ 
     HD 2770A &  9.566 &     0.61 &    -4.39 &    X &      2.8 &   5970 \\ 
     HD 2770B &  9.660 &     0.73 &    -4.39 &    X &      3.9 &   5844 \\ 
    HD 30101A &  8.782 &     0.82 &    -4.72 &    D &      1.9 &   5143 \\ 
    HD 30101B &  8.848 &     0.91 &    -4.79 &    D &      2.2 &   5061 \\ 
    HD 33334A &  8.023 &     0.70 &    -4.99 &    D &      1.9 &   5650 \\ 
    HD 33334B &  8.857 &     0.80 &    -4.63 &    U &      1.7 &   5201 \\ 
    HD 66491A &  9.253 &     0.75 &    -4.65 &    D &      2.5 &   5497 \\ 
    HD 66491B &  9.312 &     0.67 &    -4.58 &    D &      2.4 &   5492 \\ 
    HD 76037A &  7.688 &     0.50 &    -5.14 &    D &      7.9 &   6353 \\ 
    HD 76037B &  8.269 &     0.50 &    -5.03 &    D &      9.5 &   6442 \\ 
     HD 8009A &  8.819 &     0.64 &    -4.96 &    D &      0.2 &   5688 \\ 
     HD 8009B &  9.724 &     0.82 &    -4.95 &    D &      0.0 &   5291 \\ 
     HD 8071A &  7.312 &     0.57 &    -4.74 &    U &      5.5 &   6218 \\ 
     HD 8071B &  7.573 &     0.60 &    -4.71 &    U &      6.0 &   6142 \\ 
    HD 85441A &  8.907 &     0.70 &    -4.60 &    X &      1.3 &   5701 \\ 
    HD 85441B &  9.284 &     0.71 &    -4.56 &    X &      1.6 &   5537 \\ 
    HD 86057A &  8.839 &     0.60 &    -4.49 &    X &      6.0 &   6012 \\ 
    HD 86057B &  9.676 &     0.73 &    -4.40 &    X &      4.6 &   5629 \\ 
    HD 87743A &  8.734 &     0.62 &    -4.71 &    D &      2.5 &   5943 \\ 
    HD 87743B &  8.890 &     0.60 &    -4.59 &    D &      3.0 &   5905 \\ 
    HD 94399A &  9.407 &     0.61 &    -4.54 &    X &      3.2 &   5970 \\ 
    HD 94399B &  9.306 &     0.71 &    -4.56 &    X &      3.6 &   6017 \\ 
     HD 9911A &  9.428 &     0.90 &    -4.60 &    U &      1.3 &   5000 \\ 
     HD 9911B &  9.448 &     0.89 &    -4.60 &    U &      1.3 &   4968 \\ 
    HD 99121A &  8.162 &     0.46 &    -4.67 &    U &      6.7 &   6501 \\ 
    HD 99121B &  9.018 &     0.47 &    -4.57 &    U &      5.0 &   6374 \\ 
   HIP 104687A &  8.144 &     0.64 &    -4.41 &    D &      3.0 &   5870 \\ 
   HIP 104687B &  8.189 &     0.71 &    -4.48 &    D &      3.4 &   5801 \\ 
   \hline
      14 Her &  6.610 &     0.88 &    -5.06 &    D &      1.6 &   5388 \\ 
      40 Eri &  4.430 &     0.65 &    -4.90 &    D &      0.5 &   5151 \\ 
      51 Peg &  5.450 &     0.67 &    -5.08 &    D &      2.0 &   5787 \\ 
     61 Cyg B &  6.030 &     1.31 &    -4.95 &    D &      1.7 &   4077 \\ 
     83 LeoA &  6.490 &     1.00 &    -4.84 &    D &      1.4 &   5502 \\ 
     GJ 380 &  6.610 &     1.33 &    -4.72 &    D &      1.9 &   3876 \\ 
      GJ 580A &  6.580 &     0.78 &    -5.11 &    D &      2.1 &   5174 \\ 
   HD 166435 &  6.840 &     0.58 &    -4.27 &    D &      7.6 &   5964 \\ 
  $\rho$ CrB &  5.390 &     0.61 &    -5.08 &    D &      1.0 &   5823 \\ 
 $\tau$ Cet &  3.490 &     0.73 &    -4.98 &    D &      1.0 &   5283 \\ 
\hline
\end{longtable}
\end{longtab}

\begin{longtab}
\begin{longtable}{l*{5}{c}rrr}
\caption{\label{HA_RV}\textbf{SARG data:} for each star we indicate the the number of observations acquired during the SARG survey, the data (JD-2450000) of the first and the last point, the average value of \Ha, the value of \Haexcess\ and of the standard deviation of the points. In the last two columns we indicate the mean value of the RVs and its standard deviation, corrected for the known Keplerian motion, if applicable. }  \\
\hline\hline
Star & n. obs & $JD_0$& $JD_F$&$<H\alpha>$ & $\Delta H\alpha$ & $\sigma_{H\alpha}$ & RV & RMS(RV) \\
	&		&		&		&			&					&					& [km/s] & [m/s] \\
\hline
\endfirsthead
\caption{Continued.}\\
\hline\hline
Star & n. obs & $JD_0$& $JD_F$&$<H\alpha>$ & $\Delta H\alpha$ & $\sigma_{H\alpha}$ & RV & RMS(RV) \\
	&		&		&		&			&					&					& [km/s] & [m/s] \\
 \hline
\endhead
\hline 
\endfoot
   BD+182366A &     20 &    1985.4166 &    4251.3874 &    0.255 &    0.039 &    0.025 &    11.00 &    12.67 \\ 
   BD+182366B &     18 &    1985.4317 &    4251.3989 &    0.246 &    0.029 &    0.021 &    11.40 &    11.48 \\ 
   BD+222706A &     18 &    2011.6073 &    4309.4345 &    0.225 &    0.008 &    0.019 &    -4.10 &    25.41 \\ 
   BD+222706B &     18 &    2011.6264 &    4962.4638 &    0.258 &    0.045 &    0.012 &     2.41 &    21.55 \\ 
   BD+231978A &     14 &    1825.7278 &    4398.6806 &    0.371 &    0.135 &    0.027 &    20.50 &    28.41 \\ 
   BD+231978B &     13 &    1825.7152 &    4398.6929 &    0.383 &    0.149 &    0.025 &    23.50 &    35.19 \\ 
   HD 105421A &     21 &    2011.4704 &    4902.4194 &    0.237 &    0.010 &    0.017 &     7.40 &    16.87 \\ 
   HD 105421B &     19 &    2011.4840 &    4902.4333 &    0.290 &    0.069 &    0.014 &     0.44 &    12.76 \\ 
   HD 106515A &     31 &    1986.5327 &    6026.5634 &    0.214 &   -0.002 &    0.006 &     0.43 &     6.00 \\ 
   HD 106515B &     30 &    1986.5442 &    6026.5757 &    0.218 &   -0.003 &    0.007 &    18.80 &     8.69 \\ 
   HD 108421A &     17 &    1986.5975 &    4250.4681 &    0.296 &    0.044 &    0.007 &     2.00 &    21.82 \\ 
   HD 108421B &     13 &    2012.4862 &    4250.4796 &    0.360 &    0.116 &    0.018 &     2.00 &    38.30 \\ 
    HD 108574 &     22 &    1913.7615 &    4251.4385 &    0.287 &    0.064 &    0.008 &    -2.10 &    18.09 \\ 
    HD 108575 &     22 &    1913.7846 &    4251.4499 &    0.307 &    0.091 &    0.010 &    -1.50 &    33.09 \\ 
   HD 109628A &     14 &    1986.5683 &    4961.3994 &    0.214 &   -0.007 &    0.012 &     0.00 &    10.18 \\ 
   HD 109628B &     13 &    1986.5799 &    4961.4117 &    0.215 &   -0.007 &    0.010 &     0.00 &    16.16 \\ 
   HD 117963A &     15 &    2012.5413 &    5968.6519 &    0.226 &    0.003 &    0.012 &    -5.80 &    33.22 \\ 
   HD 117963B &     13 &    2012.5543 &    5968.6641 &    0.233 &    0.012 &    0.010 &     4.18 &    66.04 \\ 
   HD 118328A &     15 &    2013.6231 &    4252.5454 &    0.212 &   -0.006 &    0.013 &    19.20 &    14.38 \\ 
   HD 118328B &     14 &    2013.6353 &    4250.5043 &    0.217 &    0.001 &    0.013 &    18.40 &    15.93 \\ 
   HD 121298A &     14 &    1912.7867 &    4161.5587 &    0.229 &    0.001 &    0.009 &     0.00 &     8.28 \\ 
   HD 121298B &     12 &    1912.7733 &    4161.5702 &    0.231 &    0.006 &    0.018 &     0.00 &    12.58 \\ 
   HD 123963A &     15 &    2011.5410 &    4309.4080 &    0.222 &    0.006 &    0.011 &   -24.40 &    12.23 \\ 
   HD 123963B &     13 &    2011.5537 &    4309.4202 &    0.238 &    0.024 &    0.014 &   -24.40 &    17.28 \\ 
   HD 124054A &     13 &    2011.5702 &    4251.4849 &    0.222 &    0.002 &    0.004 &   -14.60 &     8.25 \\ 
   HD 124054B &     14 &    2011.5833 &    4251.4964 &    0.218 &    0.002 &    0.021 &   -13.40 &    10.99 \\ 
   HD 126246A &     18 &    2012.5729 &    4488.7666 &    0.343 &    0.119 &    0.008 &     0.80 &    28.88 \\ 
   HD 126246B &     16 &    2012.5846 &    4311.3850 &    0.312 &    0.092 &    0.012 &     1.70 &    14.71 \\ 
   HD 128041A &     23 &    2013.4984 &    4276.4738 &    0.210 &   -0.003 &    0.020 &   -74.70 &     7.45 \\ 
   HD 128041B &     21 &    2013.5115 &    4276.4852 &    0.230 &    0.010 &    0.034 &   -73.60 &    16.31 \\ 
   HD 132563A &     63 &    2013.6508 &    5968.6857 &    0.227 &    0.005 &    0.021 &     1.80 &    16.47 \\ 
   HD 132563B &     56 &    2013.6645 &    5968.7008 &    0.221 &    0.003 &    0.025 &     1.65 &    13.39 \\ 
   HD 132844A &     27 &    2012.6152 &    4311.4244 &    0.259 &    0.044 &    0.016 &    -3.20 &    11.87 \\ 
   HD 132844B &     26 &    2012.6027 &    4311.4359 &    0.318 &    0.103 &    0.012 &    -2.00 &    18.84 \\ 
    HD 13357A &     29 &    1801.6950 &    4849.4490 &    0.235 &    0.021 &    0.013 &    26.20 &    10.55 \\ 
    HD 13357B &     25 &    1801.7086 &    4849.4612 &    0.262 &    0.046 &    0.018 &    25.40 &    13.83 \\ 
   HD 135101A &     14 &    1982.7540 &    4488.7807 &    0.202 &   -0.011 &    0.007 &     0.00 &     4.84 \\ 
   HD 135101B &     12 &    1982.7697 &    4311.4614 &    0.212 &   -0.002 &    0.011 &     0.00 &     5.80 \\ 
   HD 139569A &     18 &    2012.6615 &    4339.4064 &    0.281 &    0.057 &    0.013 &   -29.40 &    24.53 \\ 
   HD 139569B &     20 &    2012.6733 &    4339.4179 &    0.279 &    0.062 &    0.015 &   -29.80 &    29.57 \\ 
   HD 143144A &     19 &    1798.3625 &    4339.3805 &    0.223 &    0.006 &    0.015 &   -78.50 &     9.39 \\ 
   HD 143144B &     18 &    1798.3768 &    4339.3923 &    0.221 &    0.004 &    0.017 &   -78.80 &    15.29 \\ 
   HD 146413A &     20 &    2012.6910 &    4962.5618 &    0.350 &    0.105 &    0.013 &     4.20 &     8.61 \\ 
   HD 146413B &     19 &    2012.7035 &    4962.5734 &    0.350 &    0.108 &    0.020 &     5.30 &    15.32 \\ 
    HD 17159A &     28 &    1797.6565 &    4819.3558 &    0.219 &   -0.003 &    0.011 &    11.40 &    21.43 \\ 
    HD 17159B &     28 &    1797.6727 &    4819.3679 &    0.219 &   -0.000 &    0.016 &    10.20 &    15.88 \\ 
   HD 186858A &     44 &    1798.4744 &    4962.5879 &    0.310 &    0.076 &    0.010 &    -0.63 &     8.90 \\ 
   HD 186858B &     41 &    1798.4601 &    4962.6015 &    0.297 &    0.061 &    0.013 &     1.54 &     7.64 \\ 
   HD 190042A &     23 &    1825.4814 &    4783.3459 &    0.210 &   -0.004 &    0.010 &    -4.60 &     5.77 \\ 
   HD 190042B &     22 &    1825.4615 &    4783.3593 &    0.212 &   -0.002 &    0.012 &    -3.50 &     7.74 \\ 
    HD 19440A &     19 &    1828.6588 &    4339.6564 &    0.231 &    0.005 &    0.008 &   -15.40 &    12.31 \\ 
    HD 19440B &     19 &    1828.6716 &    4339.6679 &    0.219 &   -0.002 &    0.019 &   -15.90 &     9.70 \\ 
   HD 200466A &     79 &    1801.5721 &    5807.6025 &    0.251 &    0.038 &    0.019 &    -8.00 &    15.89 \\ 
   HD 200466B &     71 &    1801.5850 &    5807.6137 &    0.247 &    0.034 &    0.014 &    -0.22 &     8.37 \\ 
   HD 201936A &     15 &    2042.6381 &    4398.3978 &    0.289 &    0.058 &    0.015 &     3.70 &    32.87 \\ 
   HD 201936B &     15 &    2042.6554 &    4398.4092 &    0.317 &    0.086 &    0.023 &     2.50 &    47.51 \\ 
   HD 209965A &     26 &    2145.5472 &    4783.3759 &    0.223 &    0.001 &    0.008 &   -19.40 &    20.60 \\ 
   HD 209965B &     22 &    2145.5634 &    4783.3878 &    0.218 &   -0.003 &    0.011 &     0.11 &    24.31 \\ 
   HD 213013A &     34 &    1827.4669 &    4369.5700 &    0.245 &    0.031 &    0.016 &   -24.70 &    10.98 \\ 
   HD 213013B &     32 &    1827.4540 &    4369.5824 &    0.264 &    0.035 &    0.024 &   -24.70 &    14.68 \\ 
   HD 215812A &     29 &    1798.4923 &    4398.4798 &    0.210 &   -0.004 &    0.005 &     8.43 &    32.68 \\ 
   HD 215812B &     18 &    1798.5063 &    4398.4912 &    0.212 &   -0.001 &    0.007 &     0.95 &    20.25 \\ 
   HD 216122A &     24 &    1801.6229 &    4962.6771 &    0.220 &    0.000 &    0.007 &   -13.30 &    18.74 \\ 
   HD 216122B &     27 &    1801.6366 &    4962.6895 &    0.225 &    0.006 &    0.012 &    -1.04 &    13.92 \\ 
   HD 219542A &     43 &    1825.5176 &    4664.6807 &    0.216 &    0.001 &    0.007 &   -12.50 &     7.43 \\ 
   HD 219542B &     48 &    1825.5048 &    4664.6931 &    0.230 &    0.016 &    0.011 &   -11.50 &     7.54 \\ 
     HD 2770A &     21 &    1856.5704 &    4338.6438 &    0.301 &    0.084 &    0.017 &    -5.00 &    21.13 \\ 
     HD 2770B &     21 &    1856.5841 &    4338.6587 &    0.313 &    0.098 &    0.023 &    -6.40 &    32.54 \\ 
    HD 30101A &     33 &    1825.6514 &    5952.4781 &    0.252 &    0.030 &    0.021 &   -18.20 &    25.26 \\ 
    HD 30101B &     33 &    1825.6652 &    5952.4943 &    0.252 &    0.027 &    0.023 &   -18.00 &    13.62 \\ 
    HD 33334A &     57 &    1801.7517 &    5952.5179 &    0.207 &   -0.007 &    0.013 &    83.20 &    22.28 \\ 
    HD 33334B &     51 &    1801.7439 &    5952.5299 &    0.216 &   -0.003 &    0.015 &    83.70 &    23.26 \\ 
    HD 66491A &     24 &    1853.7409 &    4398.7585 &    0.264 &    0.051 &    0.023 &    48.40 &    18.04 \\ 
    HD 66491B &     21 &    1853.7557 &    4161.4142 &    0.271 &    0.058 &    0.030 &    49.10 &    23.30 \\ 
    HD 76037A &     35 &    1828.7406 &    5952.6152 &    0.228 &   -0.000 &    0.012 &    22.02 &   114.52 \\ 
    HD 76037B &     34 &    1853.7833 &    5952.6280 &    0.240 &    0.009 &    0.014 &    -0.05 &    38.46 \\ 
     HD 8009A &     33 &    2116.6201 &    4819.3811 &    0.217 &    0.004 &    0.022 &   -42.10 &    12.05 \\ 
     HD 8009B &     26 &    2116.6334 &    4819.3929 &    0.223 &    0.006 &    0.020 &   -41.80 &    17.66 \\ 
     HD 8071A &     12 &    1797.6224 &    4339.6225 &    0.217 &   -0.007 &    0.005 &     5.67 &    18.94 \\ 
     HD 8071B &      9 &    1797.6397 &    3246.6791 &    0.210 &   -0.012 &    0.005 &     9.00 &    64.80 \\ 
    HD 85441A &     15 &    1826.7408 &    4754.7425 &    0.246 &    0.032 &    0.022 &   -19.80 &    12.32 \\ 
    HD 85441B &     15 &    1826.7535 &    4754.7538 &    0.264 &    0.051 &    0.016 &   -19.80 &    13.42 \\ 
    HD 86057A &     18 &    1985.5083 &    4251.3612 &    0.324 &    0.105 &    0.018 &    11.80 &    24.94 \\ 
    HD 86057B &     18 &    1985.5216 &    4251.3726 &    0.360 &    0.147 &    0.025 &    11.20 &    36.64 \\ 
    HD 87743A &     23 &    2012.3936 &    4849.6255 &    0.249 &    0.032 &    0.023 &     0.00 &    19.33 \\ 
    HD 87743B &     25 &    2012.3801 &    4849.6371 &    0.282 &    0.065 &    0.031 &     3.00 &    19.35 \\ 
    HD 94399A &     19 &    1986.4614 &    4962.3837 &    0.346 &    0.128 &    0.026 &    -6.20 &    20.16 \\ 
    HD 94399B &     17 &    1986.4731 &    4962.3953 &    0.338 &    0.119 &    0.014 &    -3.80 &    57.20 \\ 
     HD 9911A &     22 &    1801.6656 &    4339.6330 &    0.236 &    0.007 &    0.030 &   -56.60 &    11.53 \\ 
     HD 9911B &     20 &    1801.6528 &    4339.6445 &    0.223 &   -0.008 &    0.021 &   -56.30 &    11.36 \\ 
    HD 99121A &     24 &    1986.5086 &    4250.4427 &    0.223 &   -0.010 &    0.015 &    -4.40 &    23.66 \\ 
    HD 99121B &     20 &    1986.5205 &    4250.4552 &    0.217 &   -0.012 &    0.017 &    -3.10 &    30.57 \\
      HIP 104687A &     30 &    2070.6622 &    4309.6015 &    0.300 &    0.084 &    0.015 &   -20.60 &    23.82 \\ 
   HIP 104687B &     29 &    2070.6751 &    4309.6148 &    0.294 &    0.079 &    0.013 &   -21.20 &    14.20 \\  
   \hline
    14 Her &    144 &    4515.7409 &    4902.6461 &    0.223 &    0.009 &    0.003 &    -2.97 &     4.06 \\  
       40 Eri &     42 &    4515.3574 &    4819.4971 &    0.232 &    0.011 &    0.002 &   -42.20 &     7.19 \\ 
    51 Peg &     44 &    1774.6139 &    4783.4387 &    0.206 &   -0.009 &    0.003 &     0.57 &     6.00 \\ 
      61 Cyg B &    127 &    2570.3207 &    4693.6690 &    0.343 &   -0.001 &    0.008 &    -0.29 &     2.92 \\ 
      83 Leo A &    121 &    4512.5548 &    4819.6320 &    0.223 &    0.009 &    0.003 &    -2.90 &     6.54 \\ 
         GJ 380 &    145 &    4512.4711 &    4819.5728 &    0.403 &    0.013 &    0.012 &     -26.10 &    5.39 \\
   GJ 580 A &    158 &    4512.6957 &    4694.3972 &    0.213 &   -0.008 &    0.004 &   -67.90 &     7.34 \\ 
         HD 166435 &     18 &    2775.6448 &    3872.7162 &    0.411 &    0.193 &    0.010 &   -13.70 &    95.27 \\ 
      $\rho$ CrB &     46 &    2011.7355 &    4663.5609 &    0.210 &   -0.006 &    0.004 &    -1.32 &     6.09 \\
     $\tau$ Cet &    225 &    1773.7347 &    4819.3086 &    0.211 &   -0.006 &    0.003 &   -16.40 &     4.86 \\ 
 \hline
\end{longtable}
\end{longtab}

For a proper interpretation of the \Ha\ measurements that we derived 
in Sect. \ref{sec:HAindex}, some stellar parameters were considered.
We describe here the adopted sources or procedures to measure them.

Differential radial velocities were derived in \citet{CaroloPHD} and have a typical uncertainty of about 4 m/s for 
stars in the binary survey and less then 2 m/s for the bright stars. 

 We considered
measurements of \RHK\ from the literature, with preference
for studies including multi-epoch measurements to take temporal
variations of activity into account.
Overall, we retrieved \RHK\ for 36 stars from
\citet{2004ApJS..152..261W}, \citet{2010Isaacson}, 
\citet{2006A&A...454..553D},
\citet{2000A&AS..142..275S} and \citet{2003AJ....126.2048G}.
Finally, for the components of \object{HD 8009}, \object{HD 30101}, \object{HD 121298}, and \object{HD 128041}, the value of 
\RHK\ was derived from HIRES spectra available in the Keck\footnote{https://koa.ipac.caltech.edu/cgi-bin/KOA/nph-KOAlogin} archive
following the procedure described in \citet{Carolo14}.

For stars without \RHK\ values in the literature we estimated the
value from the ratio of X-ray to bolometric luminosity, using the 
calibration by \citet{2008ApJ...687.1264M}. 
This latter quantity was derived following the procedure described in
\citet{Carolo14} and \citet{CaroloPHD} for the sources identified in the ROSAT All Sky Survey 
\citep{1999A&A...349..389V,2000IAUC.7432R...1V} within 30$"$ from
our target stars. For the binaries composing most of our sample,
the components are not spatially resolved by ROSAT. We then assumed
equal X-ray luminosity for the components.
For stars that are not detected in the ROSAT All Sky Survey, this
procedure yields an upper limit on \RHK\ .
The values of \RHK\ or the upper limits, as other additional parameters we used, are listed in Table \ref{additionalparam}.

The projected rotational velocity, $v \sin i$, was obtained from a calibration
of the full with at half maximum (FWHM) of the cross-correlation function of SARG spectra.
Details will be presented elsewhere.
For the single stars we adopted the $v \sin i$ from literature sources such as
\citet{2005ApJS..159..141V}. 

The effective temperature $T_\mathrm{eff}$ of the primaries was derived from the B-V colour
using the calibration by \cite{1996Alonso} and assuming no reddening,
while for the secondaries we relied on the high-precision temperature
difference measured as part of the differential abundance analysis of 23 binary systems 
 in \cite{Desidera04,2006A&A...454..581D} and preliminary results by \cite{tesivassallo} for the others.
For the single stars (standard stars and targets of the hot-Neptune program)
we adopted the effective temperature from high-quality spectroscopic studies
\citep[e.g.][]{2005ApJS..159..141V}.

\section{\Ha\ index}
\label{sec:HAindex}

Since the Ca II H\&K lines wavelengths are not included in the SARG yellow grism spectral range, we defined a new activity index based on the \Ha\ line to study the activity of the stars in this sample.
We built an IDL procedure optimized for the SARG spectra format:
we measured the instrumental flux (not corrected for the blaze function) in a wavelength interval centred on the line core, $F_H$, and in two additional intervals symmetrically located with respect to the centre, $F_{c1}$ and $F_{c2}$.
\Ha\ is defined as $H_\alpha=2 F_H/(F_{c1}+F_{c2})$,
where $F_{c1}=flux$[6558.80\AA - 6559.80\AA], $F_H=flux$[6562.60\AA - 6563.05\AA ], and $F_{c2}=flux$[6565.20\AA - 6566.20\AA ]. 
Since the SARG spectrograph was not built to study in the  \Ha\ spectral range, this line appears twice but close to the edges of two orders (close to the blue edge of the order 93 and to the red edge of the order 94), according to the RV of the star. When we choose a wider window for $F_{c1}$ and $F_{c2}$ or increase the distance from $F_H$, the number of spectra in which the selected wavelength exits the detector therefore increases. Our choice is the best compromise.
For the same reason we  were unable to use the \Ha\ index that was used by other authors \cite[e.g.][]{2003A&A...403.1077K,2009A&A...495..959B, 2010Santos, Gomes14}: for each order one of the two continuum reference windows used by these authors is outside of the region covered by the detector. 
Furthermore, we were unable to use a reference continuum to estimate the continuum flux because it is difficult to define the proper blaze function given the presence of the extended wings of the photospheric \Ha\ absorption.
To make our measurement more reliable, we used the weighted mean of two \Ha\ values when 
fluxes for all these spectral bands could be measured in both orders.

\subsection{Error estimation}
We then analysed the possible sources of errors. 
\subsubsection*{- Internal noise} We estimated the errors on the fluxes assuming photon noise.

The error on \Ha\ index was then derived by error propagation:
 \begin{equation}
 	err_{H\alpha}=H_\alpha \sqrt{\left(\frac{1}{SN_H}\right)^2+\frac{1/c_{1err}^2+1/c_{2err}^2}{\left(F_{c1}+F_{c2}\right)^2}}
 	\end{equation}
where $SN_i=\sqrt{ gain \cdot F_i}$, $c_{1err}=SN_{c1}/F_{c1}$, $c_{2err}=SN_{c2}/F_{c2}$.
 We note that because of the lower value of the blaze echelle function, the \Ha\ indexes coming from the order 94 have a lower weight on average.
 \object{HD 128041}, \object{HD 143144}, \object{HD 9911}, \object{61 CygB} and \object{GJ 580A} have high absolute radial velocities (RV < -50 km/s) so that their spectra are remarkably blueshifted.
Their \Ha\ indexes have a  higher uncertainty because the \Ha\ line is shifted out from the order 93 spectra, therefore we were only able to use the SARG order 94, which yields poorer results.

\subsubsection*{- Systematic error}
\begin{figure}
\includegraphics[width=\columnwidth]{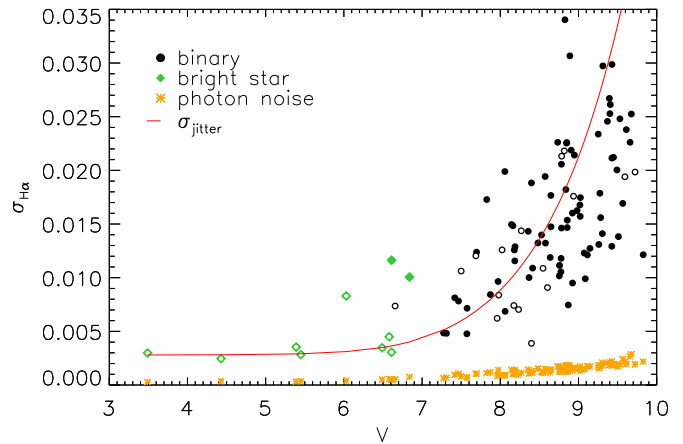}
\caption{Relation between V magnitude of the stars and the standard deviation $\sigma_{H\alpha}$ of the \Ha\ index. Open symbols indicate quiet stars (see Sect. \ref{subsec:teff} for details), green diamonds are the bright stars sample. The continuous line represents the $\sigma_{jitter}$ we adopted, while the orange asterisks indicate the mean photon noise value for each star.}
\label{V_sigma}
\end{figure}

We also considered that several other sources of noise can introduce errors on the \Ha\ index: flat fielding, background subtraction, bad pixels, instrumental instability, fringing, etc. All these contributions, added to the possible intrinsic variations of activity, increase the standard deviation of the \Ha\ values ($\sigma_{H\alpha}$).
$\tau$ Cet was used as a test target for this purpose. It is very bright and its \Haexcess\ variation is lower than 0.005 dex (peak to valley) with low levels of variability in \RHK\ from the literature. We studied the variation of $\tau$ Cet night by night. We note that the standard deviation of \Ha\ is about 10 times the intrinsic photonic error, therefore we decided to add a jitter to our measurement. 
Errors significantly larger than the photon noise error have been reported in other cases of 1 \AA\ wide activity indices from echelle spectra, see for instance \citet{2004ApJS..152..261W}.  We found that this increase does not depend on the activity level of the star. It is instead described by a relation with the stellar magnitude as shown in Fig. \ref{V_sigma}: $
\sigma_{jitter}=\sqrt{(0.0028)^2+(5.27\cdot 10^{0.4 V -6})^2}$.
Our adopted jitter is compatible with the single night variations of $\tau$ Ceti and we rescaled it for other stars according to their magnitude. 
The dependence on magnitude is that expected for error sources as background subtraction.

Finally the error applied to each measurement of \Ha\ is the sum of the photonic error and the instrumental jitter as derived above.
As the jitter is significantly larger then the photon noise, individual errors on \Ha\ index of a given star are very similar. Therefore the estimate of the jitter term has a very limited effect on the periodogram analysis presented in Sec. \ref{sec:HAtimeseries}. 

\subsubsection*{- Contamination by telluric lines}
In the spectral range of \Ha\ we considered, there are several telluric lines mainly due to the water vapour. These lines can enter in our $F_{c1}$, $F_{c2}$ and $F_H$ intervals and influence the \Ha\ index value. 
The strongest line is H$_2$O at 6564.206 \AA. If this telluric line enters $F_H$, the \Ha\ index will decrease of about 1.5\%, giving an error by about 0.005 on a quiet star. This can occur when the geocentric velocity of the star is between $52$ and $75$ km/s. Therefore only a few of our spectra are involved, but none of those discussed below.
The effect we have if this line enters $F_{c2}$ is about 0.001 dex which is negligible.
The H$_2$O line at 6560.555 \AA\ can also enter the $F_H$ interval with a comparable contribution if the geocentric radial velocities between $-90$ and $ -123$ km/s are involved. These few spectra were rejected. 
\subsubsection*{- Contamination estimation}
\begin{figure}
\includegraphics[width=\columnwidth]{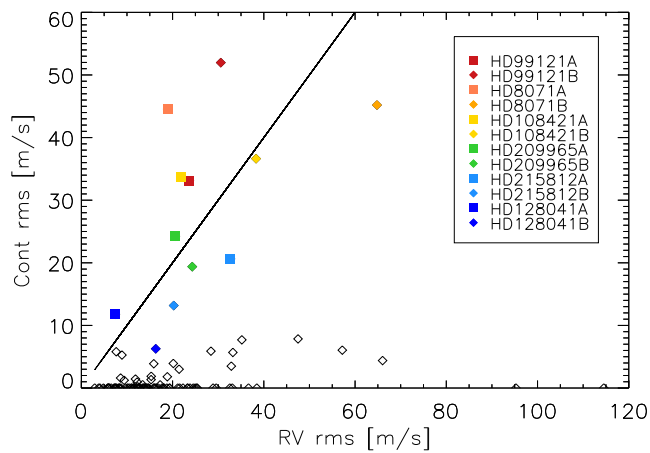}
\caption{Relation between the standard deviations of the RVs (after the correction for known Keplerian motions) and that induced by the contamination. Systems with a non-negligible contamination are highlighted.}
\label{contam}
\end{figure}

Even though during the observations of the binaries the slit was oriented perpendicularly to the separation of the components, some spectra are strongly contaminated by the companion star and were rejected. Furthermore we modelled the contamination for each consecutive observation of the companions assuming a Moffat-like shape for the point spread function (PSF) and taking into account the separation, the magnitude and the seeing. We obtain that the contribution of the contamination to the \Ha is lower then 1\% in the majority of the case and therefore is negligible. 
We found instead that for six systems the variation induced by the contamination is greater than the intrinsic variation (Fig. \ref{contam}).

For example, \object{HD 8071} is a very close binary system ($\rho = 2.183"$ according to Hipparcos) 
and the primary star is a spectroscopic binary with an amplitude of a few km/s. The effect of contamination on the RV is further modulated by the velocity of the primary at the observing epoch. This causes the RV to vary around the true value  by up to a few hundreds m/s in a quite unpredictable way. For more details see \cite{2005Martinez}.

\begin{figure}
\includegraphics[width=\columnwidth]{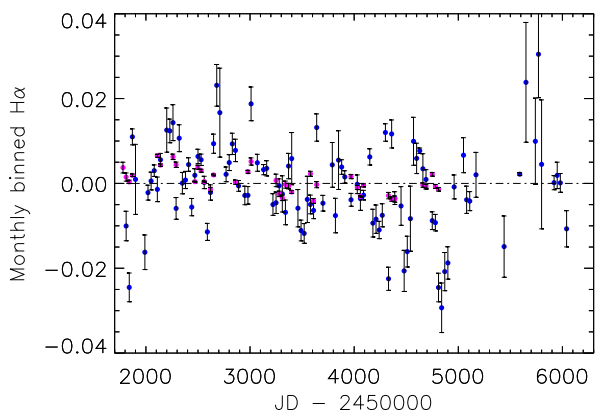}
\caption{Time evolution of the \Ha\ values of all the spectra normalised to the median value for each star, monthly bins. The red points correspond to the $\tau$ Cet data series.}
\label{stab}
\end{figure}

\begin{figure}
\centering
\includegraphics[width=\columnwidth]{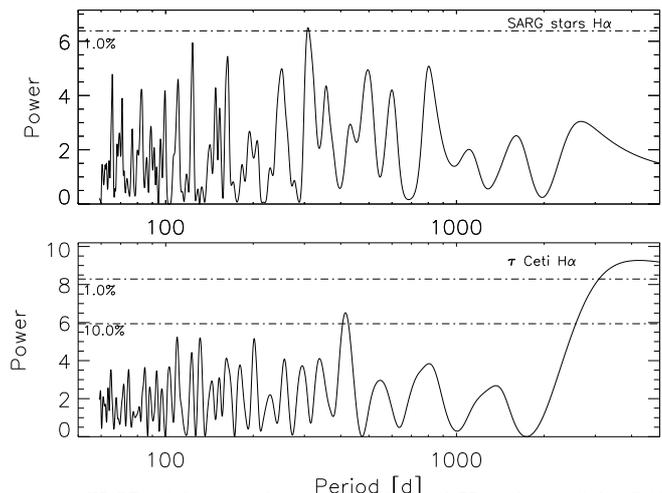} 
\caption{GLSP of the synodic monthly binned \Ha\ values of the SARG sample (top) compared to $\tau$ Cet (bottom).}
\label{medieGLSP}
\end{figure}

\subsection{Stability of the instrument}
The stability of the instrument during the survey was tested: 
for stars in the binary sample, we normalised the \Ha\ value for each spectrum to the median value of its star. 
We then binned these values into the synodic monthly mean over different stars and compared them to the same results for the $\tau$ Cet data series (see Fig. \ref{stab}). For $\tau$ Cet data we found that points are located around zero with $\sigma_{H\alpha}=0.003$. For the stars in the binary sample, the last two years of the campaign were devoted to observing mainly a few stars with candidate companions and/or RV trends, therefore the \Ha\ monthly means depend on the variability of the individual targets, as in the case of HD 200466 \citep{Carolo14}. We also verified the presence of periodicity by applying the generalize Lomb-Scargle periodogram (GLSP) \cite{GLSP} \footnote{\url{https://github.com/callumenator/idl/blob/master/Routines/Periodogram/generalised_lomb_scargle.pro}} to the two sequences of the binned values: the whole sample sequence shows no significant peak and differs from the $\tau$ Ceti sequence, which shows a long-term trend (see Fig.\ref{medieGLSP}).\\

\subsection{Dependence on T$_{eff}$ and \Haexcess\ definition}
\label{subsec:teff}
We divided our sample into two subgroups: as active stars we indicate stars with \RHK\ greater than {-4.80}, the others are called quiet stars.

\begin{figure}
 \centering
\includegraphics[width=\columnwidth]{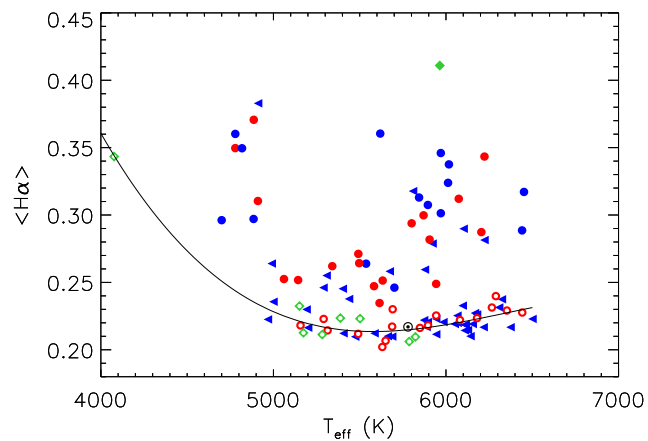}
\caption{Relation between the temperature and the median value of \Ha\ for each star. Colours are given according to $\log R_{HK}$ value sources: values from the literature are plotted in red, while for the blue dots the values are derived from X-ray luminosity. The blue triangles indicate that the \RHK\ value for a star is only an upper limit. Green diamonds indicate the bright stars sample. Open symbols correspond to quiet stars. The line only shows the fit of the binaries to have a sample unbiased by activity. The line shows the best fit for the quiet stars. The position of the Sun is also shown with $\odot$.}
\label{T_ha}
\end{figure}

Since our \Ha\ index is defined as the ratio between the flux in the line centre and the flux in the wings, we expect that different stars with the same activity level can have different \Ham\ values because of the different photospheric spectrum. Therefore we compared effective temperature 
and \Ham\, to determine the appropriate relation for quiet stars (Fig. \ref{T_ha}). 

Most of the quiet stars lie at $\langle H_\alpha \rangle\sim0.22$ for $T_{eff}>5000$ K.
At lower temperature, the \Ham\ index for quiet stars seems to increase. Active stars scatter mostly at higher \Ham.
 We also made a comparison with the Sun: it has an effective temperature of 5780 K  and its \Ham\ is 0.217, as measured in the solar flux atlas \citep{1984sfat.book.....K}, in agreement with the lower envelope for quiet stars.
We describe the distribution of the quiet stars in this lower envelope with a three-degree function 
 and define as \Ha-excess (\Haexcess) the point distance from this line:  \Haexcess\ is the difference in \Ha\ index of a star with respect to a quiet star that has the same effective temperature. 
Therefore we decided to use \Haexcess\ as the activity index; it is more robust than \Ha\ because it allows us to compare the activity of stars with different temperatures. 
  
 \section{Sample analysis} 
\label{sec:sampleanalysis}

\subsection{Correlation with \RHK and rotation}
\begin{figure}
\includegraphics[width=\columnwidth]{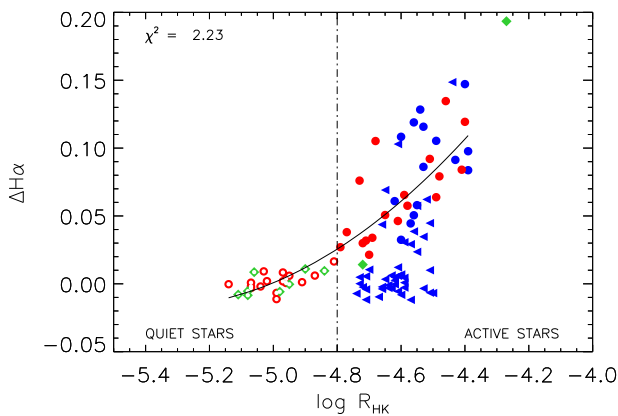}
\caption{Relation between the \RHK\ and \Haexcess. Colours are the same as Fig. \ref{T_ha}. Bright stars and stars with upper limits for \RHK were not considered to have a sample unbiased by activity.}
\label{RHK_Haex}
\end{figure}
\Haexcess\  correlates quite well with \RHK\ (reduced $\chi^2=2.26$, Fig. \ref{RHK_Haex}). Active stars are more scattered but typically show excess in the \Ha\ index (\Haexcess $> 0$).
All the stars for which \RHK\ has been derived from the X-ray luminosity are in the active portion of the diagram. This is due to the flux limit of the ROSAT All Sky Survey, which is only sensitive to the active stars at the typical distance of our program stars. The stars for which only upper limits are derived populate the lower envelope of the distribution in most cases: this is consistent with a low activity level.

This new index appears to show that stars are distributed in two groups, which suggests the presence of the Vaughan-Preston gap at $\Delta H\alpha=0.02$ \citep{1980PASP...92..385V}.

The results of \cite{2009A&A...499L...9P} also show the presence of a gap between $\log R_{HK} = -4.7 $ and $-5.0$. This corresponds to the interval $\Delta H_\alpha \sim [0.01, 0.04]$.

We also found a weak relation between \Haexcess\ and its standard deviation:

the intrinsic variation of  the \Ha\ index and internal errors contribute to the increase in scatter in the \Ha\ index measurement for each star, but since the scatter is dominated by intrinsic errors for fainter stars, only the deviation seen in brighter stars is dominated by the intrinsic variability.

We also checked the well-known relation between rotation and activity \citep[e.g.][]{1984Noyes,1985ARA&ABaliunas, 2000A&A...361..265S}

We found, as expected, that a moderate rotation is enough to cause a high activity for cold stars and in this case the $v \sin i$ value increases with activity, 
while the hottest stars only have high activity values if $v\sin i$ is high: this behaviour can be related to the decrease in thickness of the convective envelope as the stars become hotter \citep{Charbonneau2012}.   

\subsection{Binary components}
\begin{figure}
\includegraphics[width=\columnwidth]{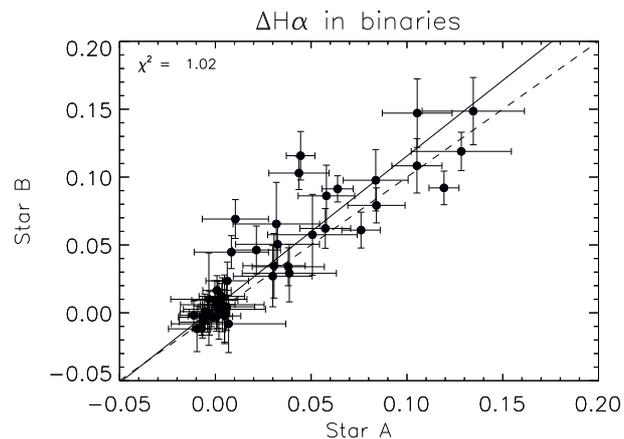}
\caption{Relation between the \Haexcess\ of the two companion stars. The solid line corresponds to the best fit, the dashed line corresponds to the equivalence.}
\label{HAab}
\end{figure}
We can compare the \Haexcess\ index for the two components in each binary system: we find a very good relation between the two stars indexes, that is
$
\Delta H\alpha_B =(1.11\pm 0.08)\Delta H\alpha_A+0.004\pm0.004$, as shown in Fig. \ref{HAab}. 
The value of the reduced $\chi^2$ suggests that the scatter is dominated by the measurement error. We tested that the long-scale activity cycles (like the solar cycle) induce a variation in \Ha\ that is weaker than our adopted measurement error.
HD 108421, HD 132844 and HD 105421 lie above the relation, but we did not note any evidence of errors in our analysis for these stars, so that the discrepancy seems to be real 
and the two stars of these systems could be in different activity phases. 
For HD 126246, which lies below the relation, the difference in the \Ha\ activity level between the two components  qualitatively agrees with the  \RHK\ and $v \sin i$ difference found by \cite{2006A&A...454..581D}, supporting an intrinsic rotation and activity difference between the two components.

\subsection{Age-activity relation}
Prompted by this result, we tested whether our \Ha\ could be an age indicator for these stars  
 \citep[e.g.][]{1972ApJSkumanich,1985ARA&ABaliunas, 1993ApJS...85..315S,Mamajek2008, 2013A&APace, 2011AJZhao}.
 We computed the ages of the binary systems with the isochrone fitting algorithm developed by \cite{2015A&ABonfanti}. The implementations details can be found in \cite{2015A&ABonfanti,2016A&ABonfanti}. Here we recall that it enables recovering the isochronal age of a field star when at least its [Fe/H], $T_\mathrm{eff}$ and $\log g$ are available. In our case we also considered \RHK as input parameter, which allowed us to disregard unlikely very young isochrones, so that we could better constrain the stellar age. Since the evolution of low-mass stars is extremely slow, this method works well for stars with $T_\mathrm{eff}$>5500 K; for cooler (less massive) stars, uncertainties in the exact location of a star on the Hertzspurng-Russel (HR) diagram leads to an error so large that practically all ages from 0 up to the age of the Universe are possible. We therefore did not consider such stars in our test.

From the differential abundance analysis, $T_\mathrm{eff}$ and $\log g$ have typical uncertainties of $\sim50$ K and $\sim0.15$ dex, respectively, while the differences $\Delta T_{eff}=T_{effA}-T_{effB}$ and $\Delta \log g=\log g_A -\log g_B$ are more reliable and their reference uncertainties have been estimated in $\sim20$ K and $\sim 0.06$ dex, respectively. 
We therefore constructed a grid in $T_\mathrm{eff}$ and $\log g$ for each binary component, with step sizes of 25 K and 0.05 dex, respectively. We discarded all the points in the grid where the relations $\Delta T_{eff}- \delta T_{eff B} < \vert T_{eff A}-T_{eff B} \vert < \Delta T_{eff}+ \delta T_{eff}$ and 
$\Delta \log g- \delta \log g < | \log g_A-\log g_B | < \Delta \log g +\delta \log g$  were not satisfied. We computed the ages of each component for each remaining point in the grid and retained only those for which the stars could be considered coeval ($\vert \log t_A - \log t_B \vert <0.05$; 0.05 is the resolution of the isochrone grids).
For each analysed star, we built a catalogue reporting the plausible input parameters and the resulting age that was coeval to that of its companion. For each binary system, we synthesised these data providing the youngest and oldest feasible age of the system and the median age.
\begin{figure}
\centering
\includegraphics[width=1.05\columnwidth]{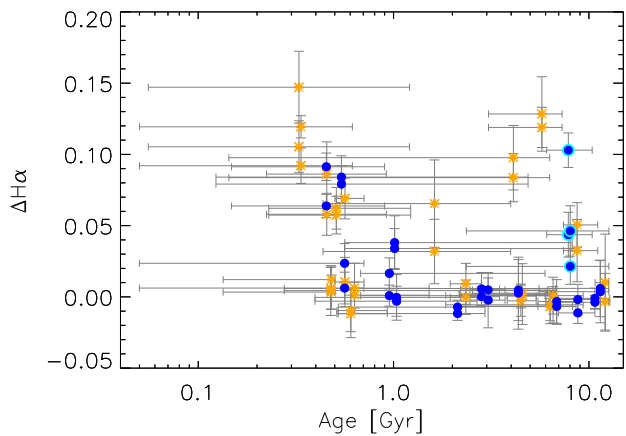}
\caption{Activity as a function of the age. Blue circles represent the star in \cite{Desidera04} for which we have solid constraints on the temperature; orange crosses show the other stars. The two systems with an uncertain parallax are highlighted in cyan. }
\label{age}
\end{figure}

In Fig. \ref{age} we plot for each star hotter than $T_{eff} = 5500$ K its \Haexcess\ as a function of the age of the system. We divided the systems into two subsamples according to the reliability of the input parameters, and in particular the $T_\mathrm{eff}$: blue dots represent the systems analysed in \cite{Desidera04}, which are more accurate, while orange crosses correspond to preliminary results for systems analysed in \cite{tesivassallo}.
The result shows that the majority of the active star are younger than 1.5 Gyr, while for older stars the distribution is flattened around zero, that is, they are inactive.\\
We found that the activity for young stars  is anti-correlated with the age, confirming that the relation between the \Haexcess\ in the components of the systems younger than 1.5 Gyr is mainly due to age. 
The position of the pairs HD132844A and B and HD13357A and B in the diagram of Fig. \ref{age} does not follow the general trend: the position on colour-diagram of HD132844 below the main sequence \citep[see][]{Desidera04} is indicative of substantial error in the trigonometric parallax.
The two Hipparcos solutions for the parallax of HD13357 are inconsistent with each other.
In both cases we can conclude that there is an underestimated error in the parallax.
Indeed, the adopted parameters (especially the gravity) depend on the adopted trigonometric parallax: 
in the abundance analysis the effective temperatures were derived from ionization equilibrium and stellar gravities from luminosities, masses and temperatures, using iterative procedures.
It seems therefore that a well-defined activity-age relation persists only for objects younger than $\sim 1.5$ Gyr, and  that after this age \Ha\ seems to be less efficient as an age indicator. Our data did not show significant correlation between these quantities: due to the lack of data with such an age, we cannot conclude whether if there is a discontinuity or if the activity of the star decreases with time. The activity-age anti-correlation for younger stars confirms results from \cite{1988ApJBarry}, for example,  and the apparent flatness of the plot for older stars seems to agree with \cite{2013A&APace};but owing to the uncertainty on our ages, we cannot confirm or reject the idea that the activity decreases with age also for older stars, with a different slope as found by \cite{Mamajek2008}, for instance. Finally, we found that a large portion (15 over 35) of the stars in our sample with age estimates from the isochrone method are younger than 1.5 Gyr: this could be due to the recent bump in the star formation rate in the solar neighbourhood as claimed by \cite{1988ApJBarry} or to a bias in the age distribution of the stars in the Hipparcos Multiple Stellar Catalog. Future observations of results from the GAIA satellite may clarify this question.\\

\subsection{Activity vs RV scatter}
We finally found the well-known relation between the activity of a star and the standard deviation of its radial velocities \citep[]{1997ApJ...485..319S, 1998ApJ...498L.153S, 2000A&A...361..265S, 2009A&A...495..959B, 2011A&A...528A...4B}. In addition, when considering the contamination of the spectra, we found that it is not negligible especially for the systems HD 8071, HD 99121, HD 108421 and HD 209965, which were omitted in this discussion and are detailed below. There are also a number of cases for which the spread in RV during the survey is high ($ > 80 m/s$) and which have a relatively low activity level.

\begin{figure}
\includegraphics[width=\columnwidth]{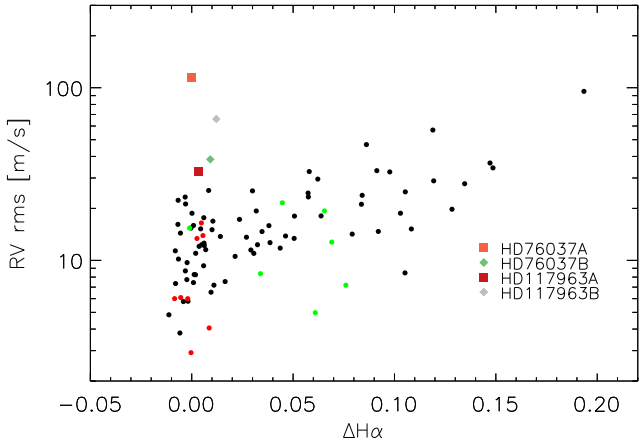}
\caption{Relation between the \Haexcess\ and the RV standard deviation in the survey. The green dots indicate the RV standard deviation of the stars with a known Keplerian trend that is due to a companion, in red we plot the RV standard deviation of stars with a known companion. In both cases we correct the data for their known RV variation.}
\label{HAex_RVrms}
\end{figure}
Most of these objects have known RV trend of Keplerian origin and after the RV variation induced by the companion was removed, they became part of the main trend (Fig. \ref{HAex_RVrms}).
In addition there are at least four stars left outside the general trend. Since these are potentially very interesting objects, we examine them more in detail.

\object{HD 76037A} and B: this is a wide binary composed of two F-type stars. The SARG spectra show that the primary star is a long-period partially blended SB2 star, therefore we conclude that the excess scatter in RVs is due to the blending of the spectra of the two components. For the secondary, the excess of the RV scatter is fairly large even after resuming the long-term trend with time that indicates the presence of a low-mass companion; in addition, the \Ha index also has a trend with time - more likely related to a cycle.

The Hipparcos Multiple Stellar Catalog indicates that the  \object{HD 117963A} system has a separation wide enough to rule out contamination effects ($\rho=3.493"$).
\object{HD 117963B} is a spectroscopic binary and some of the spectra were taken with low S/N (Desidera et al. in prep.). 

We cannot exclude Keplerian motion as the origin of the scatter for both stars, therefore a deeper analysis with acquisition of additional data would be required.

\section{H$\alpha$ index time-series analysis}
\label{sec:HAtimeseries}

By analogy with the Sun, emission in the core of \Ha\ is expected to show time variability mainly modulated by
stellar rotation over a period of the order of days, and by the activity cycle over periods of hundreds or thousands  of days. 
In addition, secular variations in the activity levels similar to the Maunder minimum can be present.
Therefore the different properties of the time series of our objects should be taken into account. Stars in the Hot-Neptunes program were observed for a single season with a moderately dense sampling. 
In this case rotation periods could be found, but periodicities due to the activity cycle cannot be reliably identified. On the other hand, for the SARG survey objects, the observational campaign was longer and less dense. 
For only a few targets do we have a larger number of spectra because during the survey they were suspected to host a planet. This was the case of \object{HD 106515A} \citep{Desidera12} and \object{HD 132563B} \citep{Desidera11}, for example. In addition we already know that for \object{HD 200466A}, the RV variations seen are mainly due to an activity cycle \citep{Carolo14}.

It is known that more active stars have irregular periods that are not easy to 
determine with the analysis of periodograms. In spite of this, we computed the GLSPs for the \Ha\ index that was obtained using the \cite{GLSP} procedure. 
To evaluate the significance of these periodicities, the false alarm probability (FAP) of the highest peak of the periodogram was estimated through a bootstrap method, with 1000 permutations. We used the spectral window function to rule out that our periodicity is due to the sampling.
The results for the most interesting objects are listed in Table \ref{HAresults}. 

We found a signature of periodic variations (rotational periods or activity cycles) in 19 stars, whereas 10 stars show a clear overall trend in \Ha\ with time. 
On the other hand the stars for which we were able to find evidence of activity cycles are all with moderate activity excess and temperatures of between 4800 and 6000 K. The stars showing a long-term trend are hotter than average.

It is noteworthy that of the binary stars that show promising cycles, only \object{HD 76037A\&B} are quiet and show a long-term trend.\\

Of the bright stars,
\object{51 Peg} was used as a RV standard to monitor instrument performances during the binary program. The quite good temporal coverage of the data allowed us to detect a significant long-term period of about seven years with FAP of 0.6\%. Added to this signal, we also found a periodicity of 86.49 d, which corresponds to an alias of the $21.9\pm 0.4$ d period found by \cite{2010MNRAS.408.1666S} with one sinodic month.
This shortest period seems then to be the rotational signal. We obtained a similar result also for \object{61 Cyg b}: the GLSP peaks at 16.44 d, which is an alias of the $\sim 37$ d period 
\citep{2007ApJ...657..486B, 2009A&A...501..703O}. \object{14 Her} shows a periodicity of 22.38 d. In this case the
spectral window is complex and we cannot rule out that this period is fake. \cite{2004ApJS..152..261W} estimated a rotational period of 48 days from the \RHK\ mean value, but this was not detected by \cite{2010MNRAS.408.1666S}. \\

All the \Ha\ time series are presented in Table 5, only available in electronic form
at the CDS.

\begin{table*}
\centering
\begin{tabular}{lcrrrrr}
    \hline
    \hline
Star & & \Haexcess &  Rotation & Cycle & Amplitude & FAP\\
 & & & [d] & [d] & &[\%] \\
\hline

\hline
 HD 186858 & A     & 0.077  & 7.68  & --&0.010& 1.0 \\
      & B     & 0.062  & --    & 2030 & 0.014& 0.3\\
\hline
HD 200466 & A     & 0.038  & --    & 1500 &0.024& $<0.1$\\

      & B     & 0.034  & --    & trend &>0.015 & $<0.1$  \\
\hline
BD+182366 & A     & 0.039  & --    & 1432&0.037  & 0.1\\

      & B     & 0.029  & --    & -- & --&\\
\hline
HD 139569 & A     & 0.057  & --    & trend & > 0.21 &2.2 \\

      & B     & 0.062  & --    & trend & >0.30 & 0.6\\
\hline
HD 76037 & A     & -0.002  & --    & trend & >0.014 &1.0\\
      & B     & 0.014  & --  & trend & >0.022 &1.3\\
\hline
HD 201936 & A     & 0.059  & 13.70  & -- &0.020& 0.7 \\

      & B     & 0.087  & --    & -- & \\
\hline
HD 213013 & A     & 0.031  & --    & --& --\\
& B     & 0.035  & 3.59  & --& 0.019 &2.1\\
\hline
14 Her & & 0.009  & 22.38 &-- &0.002 &$<0.1$\\
\hline
51 Peg & & 0.003 & 86.49 & 2069 & 0.001, 0.003& 0.8, 0.6\\
\hline
61 Cyg B & &-0.001  &16.44 & --& 0.011&$<0.1$\\
\hline
GJ 380 & & 0.013  &-- & trend & >0.017& $<0.1$\\
\hline
$\tau$ Ceti & & -0.006 & -- & trend & >0.003& $<0.1$\\
\hline
\end{tabular}
\caption{Stars with cycles. In the second column we indicate the component, Col. 3 reports the \Haexcess\ for the stars, Col. 4 is the short period, compatible with the rotation in our analysis (where available), Col. 5 reports the period or long-term activity cycle. Column 6 shows the amplitude of the \Ha\ variation. The last column indicates the false-alarm probability related to the identified periods.}
\label{HAresults}
\end{table*}

\section{Correlation between RV and \Ha}
\label{sec:correlation}

The high uncertainty on the single measurements of \Ha\ prevent us from properly studying the correlation with the RVs. However this was possible in some particular case, such as spectra with high S/N or stars with a relevant trend in \Ha.
We used the Spearman correlation coefficient $\rho_S$ and its significance \sig\ to quantify the correlation between RV and \Ha\ index (Table \ref{tab:correlation}): we obtained an extremely high significance for \object{HD 200466A} \citep[see]{Carolo14}. For four other objects, the probability that the correlation is the result of a random effect is lower than 0.0075.
\object{HD 201936A} and \object{HD 213013A} are active stars with a signature of an activity cycle, \object{GJ 380} spectra have a high S/N and show a probable long-term cycle. Plots are presented in the Appendix.  
In \object{HD 76037A} the anti-correlation simply shows that
both RV and activity are time-dependent on long scales. We can
therefore rule out a strong physical connection
between these two quantities for this star.

\begin{table}
\centering
\begin{tabular}{lrrr}
    \hline
    \hline
Star      & $\rho_s$ & $\sigma$ & $n_\sigma$ \\
\hline
HD 200466A & 0.556 & 0.000 & -4.817 \\
HD 76037A & -0.665 & 0.000 & 3.702 \\
HD 99121A & 0.6511 &	0.002	& -2.84 \\
HD 213013A & 0.478 & 0.008 & -2.576 \\
GJ 380     & 0.535 & 0.018 & -2.270 \\
\hline
\end{tabular}
\caption{Rank of the Spearman correlation coefficient  $\rho_s$ and its significance between \Ha\ and RV for the stars of the sample. Column 3 reports its false-alarm probability and the last column reports the $n_\sigma$ value. Only stars with significance < 0.02 are indicated. }
\label{tab:correlation}
\end{table}

\section{Conclusions}
\label{sec:conclusions}

The activity of 104 stars observed with the SARG spectrograph was studied using an index based on the \Ha\ line.
We found that this index, \Haexcess, correlates well with the index based on Ca II lines, \RHK, and therefore it can be used to estimate the average activity level, confirming previous results.   
It also correlates with the rotation of the star: low activity corresponds to slow rotation, especially for cool stars. 
After removing a few targets for which contamination of the spectra by their companion is the dominant source of RV scatter, we found that \Haexcess\ also correlates with the scatter in RV. We obtain that a low-mass companion might be the source of a high residual RV scatter at least for \object{HD 76073B}.
We also found a strong correlation between the average activity level \Ham\ of the two 
components in each binary system  and that roughly a half of our systems are active. Finally, we showed that activity as measured by \Haexcess\ is correlated with the age derived from isochrone fitting.  Although these have large error bars due to uncertainties in temperature and parallaxes, we found that active stars are typically younger than 1.5 Gyr, while older stars are typically inactive.

We then analysed the time series of the stars: 11 stars ($\sim 8.5$ \%) of the SARG sample  show a periodicity in \Ha with false-alarm probability $<0.5\%$. 
 All these stars have a moderate activity level ($0.029 < \Delta H\alpha < 0.077$) except for the pair HD 76037A and B,  but in these cases we only have a hint of a long-term period or magnetic cycle. When we focused on the long-term cycle, we obtained that the temperature interval of these stars is also limited to late-G and early-K stars. Other stars show variabilities on temporal scales certainly different from the rotational periods.
In the bright stars sample, we found five stars out of ten with significant periodic variations in \Ha. 
In some cases the physical origin of this type of signal is unclear. 

Only five stars show a significant correlation between \Ha\ and RVs. 

We conclude that if care is exerted, \Ha\  is  a useful indicator for activity and can be a good alternative to Ca II \RHK\ for studies based on radial velocity techniques, especially for solar-type stars.\\

\textit{Acknowledgements}. This research has made use of the SIMBAD database, operated
at CDS, Strasbourg, France. 
This research has made use of the Keck Observatory Archive (KOA), which is operated by the W.M. Keck Observatory and the NASA Exoplanet Science Institute (NExScI), under contract with the National Aeronautics and Space Administration. 
We thank the TNG staff for contributing to the observations and the TNG TAC for the generous allocation of observing time. 
This work was partially funded by PRIN-INAF 2008 “Environmental effects in the formation and evolution of extrasolar planetary systems”.

\bibliographystyle{aa} 
\bibliography{biblio} 

\begin{thebibliography}{59}
\expandafter\ifx\csname natexlab\endcsname\relax\def\natexlab#1{#1}\fi

\bibitem[{{Alonso} {et~al.}(1996){Alonso}, {Arribas}, \&
  {Martinez-Roger}}]{1996Alonso}
{Alonso}, A., {Arribas}, S., \& {Martinez-Roger}, C. 1996, \aap, 313, 873

\bibitem[{{Baliunas} \& {Vaughan}(1985)}]{1985ARA&ABaliunas}
{Baliunas}, S.~L. \& {Vaughan}, A.~H. 1985, \araa, 23, 379

\bibitem[{{Barry}(1988)}]{1988ApJBarry}
{Barry}, D.~C. 1988, \apj, 334, 436

\bibitem[{{B{\"o}hm-Vitense}(2007)}]{2007ApJ...657..486B}
{B{\"o}hm-Vitense}, E. 2007, \apj, 657, 486

\bibitem[{{Boisse} {et~al.}(2011){Boisse}, {Bouchy}, {H{\'e}brard}, {Bonfils},
  {Santos}, \& {Vauclair}}]{2011A&A...528A...4B}
{Boisse}, I., {Bouchy}, F., {H{\'e}brard}, G., {et~al.} 2011, \aap, 528, A4

\bibitem[{{Boisse} {et~al.}(2009){Boisse}, {Moutou}, {Vidal-Madjar}, {Bouchy},
  {Pont}, {H{\'e}brard}, {Bonfils}, {Croll}, {Delfosse}, {Desort}, {Forveille},
  {Lagrange}, {Loeillet}, {Lovis}, {Matthews}, {Mayor}, {Pepe}, {Perrier},
  {Queloz}, {Rowe}, {Santos}, {S{\'e}gransan}, \& {Udry}}]{2009A&A...495..959B}
{Boisse}, I., {Moutou}, C., {Vidal-Madjar}, A., {et~al.} 2009, \aap, 495, 959

\bibitem[{{Bonfanti} {et~al.}(2016){Bonfanti}, {Ortolani}, \&
  {Nascimbeni}}]{2016A&ABonfanti}
{Bonfanti}, A., {Ortolani}, S., \& {Nascimbeni}, V. 2016, \aap, 585, A5

\bibitem[{{Bonfanti} {et~al.}(2015){Bonfanti}, {Ortolani}, {Piotto}, \&
  {Nascimbeni}}]{2015A&ABonfanti}
{Bonfanti}, A., {Ortolani}, S., {Piotto}, G., \& {Nascimbeni}, V. 2015, \aap,
  575, A18

\bibitem[{Carolo(2012)}]{CaroloPHD}
Carolo, E. 2012, PhD thesis, University of Padua, Italy

\bibitem[{{Carolo} {et~al.}(2014){Carolo}, {Desidera}, {Gratton}, {Martinez
  Fiorenzano}, {Marzari}, {Endl}, {Mesa}, {Barbieri}, {Cecconi}, {Claudi},
  {Cosentino}, \& {Scuderi}}]{Carolo14}
{Carolo}, E., {Desidera}, S., {Gratton}, R., {et~al.} 2014, \aap, 567, A48

\bibitem[{{Charbonneau} \& {Steiner}(2012)}]{Charbonneau2012}
{Charbonneau}, P. \& {Steiner}, O. 2012, Solar and Stellar Dynamos: Saas-Fee
  Advanced Course 39 Swiss Society for Astrophysics and Astronomy, Saas-Fee
  Advanced Course (Springer Berlin Heidelberg)

\bibitem[{{Cincunegui} {et~al.}(2007){Cincunegui}, {D{\'{\i}}az}, \&
  {Mauas}}]{2007Cincunegui}
{Cincunegui}, C., {D{\'{\i}}az}, R.~F., \& {Mauas}, P.~J.~D. 2007, \aap, 469,
  309

\bibitem[{{Desidera} {et~al.}(2011){Desidera}, {Carolo}, {Gratton}, {Martinez
  Fiorenzano}, {Endl}, {Mesa}, {Barbieri}, {Bonavita}, {Cecconi}, {Claudi},
  {Cosentino}, {Marzari}, \& {Scuderi}}]{Desidera11}
{Desidera}, S., {Carolo}, E., {Gratton}, R., {et~al.} 2011, \aap, 533, A90

\bibitem[{{Desidera} {et~al.}(2012){Desidera}, {Gratton}, {Carolo}, {Martinez
  Fiorenzano}, {Endl}, {Mesa}, {Cecconi}, {Claudi}, {Cosentino}, {Scuderi},
  {Sozzetti}, \& {Zurlo}}]{Desidera12}
{Desidera}, S., {Gratton}, R., {Carolo}, E., {et~al.} 2012, \aap, 546, A108

\bibitem[{{Desidera} {et~al.}(2007){Desidera}, {Gratton}, {Endl}, {Martinez
  Fiorenzano}, {Barbieri}, {Claudi}, {Cosentino}, {Scuderi}, \&
  {Bonavita}}]{Desidera2007}
{Desidera}, S., {Gratton}, R., {Endl}, M., {et~al.} 2007, ArXiv e-prints

\bibitem[{{Desidera} {et~al.}(2006{\natexlab{a}}){Desidera}, {Gratton},
  {Lucatello}, \& {Claudi}}]{2006A&A...454..581D}
{Desidera}, S., {Gratton}, R.~G., {Lucatello}, S., \& {Claudi}, R.~U.
  2006{\natexlab{a}}, \aap, 454, 581

\bibitem[{{Desidera} {et~al.}(2006{\natexlab{b}}){Desidera}, {Gratton},
  {Lucatello}, {Claudi}, \& {Dall}}]{2006A&A...454..553D}
{Desidera}, S., {Gratton}, R.~G., {Lucatello}, S., {Claudi}, R.~U., \& {Dall},
  T.~H. 2006{\natexlab{b}}, \aap, 454, 553

\bibitem[{{Desidera} {et~al.}(2004){Desidera}, {Gratton}, {Scuderi}, {Claudi},
  {Cosentino}, {Barbieri}, {Bonanno}, {Carretta}, {Endl}, {Lucatello},
  {Martinez Fiorenzano}, \& {Marzari}}]{Desidera04}
{Desidera}, S., {Gratton}, R.~G., {Scuderi}, S., {et~al.} 2004, \aap, 420, 683

\bibitem[{{Dumusque} {et~al.}(2011){Dumusque}, {Lovis}, {S{\'e}gransan},
  {Mayor}, {Udry}, {Benz}, {Bouchy}, {Lo Curto}, {Mordasini}, {Pepe}, {Queloz},
  {Santos}, \& {Naef}}]{Dumusque11}
{Dumusque}, X., {Lovis}, C., {S{\'e}gransan}, D., {et~al.} 2011, \aap, 535, A55

\bibitem[{{Endl} {et~al.}(2000){Endl}, {K{\"u}rster}, \& {Els}}]{Endl2000}
{Endl}, M., {K{\"u}rster}, M., \& {Els}, S. 2000, \aap, 362, 585

\bibitem[{{Gomes da Silva} {et~al.}(2014){Gomes da Silva}, {Santos}, {Boisse},
  {Dumusque}, \& {Lovis}}]{Gomes14}
{Gomes da Silva}, J., {Santos}, N.~C., {Boisse}, I., {Dumusque}, X., \&
  {Lovis}, C. 2014, \aap, 566, A66

\bibitem[{{Gomes da Silva} {et~al.}(2011){Gomes da Silva}, {Santos}, {Bonfils},
  {Delfosse}, {Forveille}, \& {Udry}}]{Gomes11}
{Gomes da Silva}, J., {Santos}, N.~C., {Bonfils}, X., {et~al.} 2011, \aap, 534,
  A30

\bibitem[{{Gratton} {et~al.}(2009){Gratton}, {Desidera}, \&
  {Claudi}}]{GrattonHotNeptunes}
{Gratton}, R., {Desidera}, S., \& {Claudi}, R. 2009, MSAIt, 80, 312

\bibitem[{{Gratton} {et~al.}(2001){Gratton}, {Bonanno}, {Bruno}, {Cali},
  {Claudi}, {Cosentino}, {Desidera}, {Diego}, {Farisato}, {Martorana},
  {Rebeschini}, \& {Scuderi}}]{Gratton01}
{Gratton}, R.~G., {Bonanno}, G., {Bruno}, P., {et~al.} 2001, Experimental
  Astronomy, 12, 107

\bibitem[{{Gray} {et~al.}(2003){Gray}, {Corbally}, {Garrison}, {McFadden}, \&
  {Robinson}}]{2003AJ....126.2048G}
{Gray}, R.~O., {Corbally}, C.~J., {Garrison}, R.~F., {McFadden}, M.~T., \&
  {Robinson}, P.~E. 2003, \aj, 126, 2048

\bibitem[{{Isaacson} \& {Fischer}(2010)}]{2010Isaacson}
{Isaacson}, H. \& {Fischer}, D. 2010, \apj, 725, 875

\bibitem[{{K{\"u}rster} {et~al.}(2003){K{\"u}rster}, {Endl}, {Rouesnel}, {Els},
  {Kaufer}, {Brillant}, {Hatzes}, {Saar}, \& {Cochran}}]{2003A&A...403.1077K}
{K{\"u}rster}, M., {Endl}, M., {Rouesnel}, F., {et~al.} 2003, \aap, 403, 1077

\bibitem[{{Kurucz} {et~al.}(1984){Kurucz}, {Furenlid}, {Brault}, \&
  {Testerman}}]{1984sfat.book.....K}
{Kurucz}, R.~L., {Furenlid}, I., {Brault}, J., \& {Testerman}, L. 1984, {Solar
  flux atlas from 296 to 1300 nm}

\bibitem[{{Lovis} {et~al.}(2011){Lovis}, {Dumusque}, {Santos}, {Bouchy},
  {Mayor}, {Pepe}, {Queloz}, {S{\'e}gransan}, \& {Udry}}]{Lovis11}
{Lovis}, C., {Dumusque}, X., {Santos}, N.~C., {et~al.} 2011, ArXiv e-prints

\bibitem[{{Mamajek} \& {Hillenbrand}(2008{\natexlab{a}})}]{2008ApJ...687.1264M}
{Mamajek}, E.~E. \& {Hillenbrand}, L.~A. 2008{\natexlab{a}}, \apj, 687, 1264

\bibitem[{{Mamajek} \& {Hillenbrand}(2008{\natexlab{b}})}]{Mamajek2008}
{Mamajek}, E.~E. \& {Hillenbrand}, L.~A. 2008{\natexlab{b}}, \apj, 687, 1264

\bibitem[{{Mart{\'{\i}}nez Fiorenzano} {et~al.}(2005){Mart{\'{\i}}nez
  Fiorenzano}, {Gratton}, {Desidera}, {Cosentino}, \& {Endl}}]{2005Martinez}
{Mart{\'{\i}}nez Fiorenzano}, A.~F., {Gratton}, R.~G., {Desidera}, S.,
  {Cosentino}, R., \& {Endl}, M. 2005, \aap, 442, 775

\bibitem[{{Meunier} \& {Delfosse}(2009)}]{2009Meunier}
{Meunier}, N. \& {Delfosse}, X. 2009, \aap, 501, 1103

\bibitem[{{Noyes} {et~al.}(1984){Noyes}, {Hartmann}, {Baliunas}, {Duncan}, \&
  {Vaughan}}]{1984Noyes}
{Noyes}, R.~W., {Hartmann}, L.~W., {Baliunas}, S.~L., {Duncan}, D.~K., \&
  {Vaughan}, A.~H. 1984, \apj, 279, 763

\bibitem[{{Ol{\'a}h} {et~al.}(2009){Ol{\'a}h}, {Koll{\'a}th}, {Granzer},
  {Strassmeier}, {Lanza}, {J{\"a}rvinen}, {Korhonen}, {Baliunas}, {Soon},
  {Messina}, \& {Cutispoto}}]{2009A&A...501..703O}
{Ol{\'a}h}, K., {Koll{\'a}th}, Z., {Granzer}, T., {et~al.} 2009, \aap, 501, 703

\bibitem[{{Pace}(2013)}]{2013A&APace}
{Pace}, G. 2013, \aap, 551, L8

\bibitem[{{Pace} {et~al.}(2009){Pace}, {Melendez}, {Pasquini}, {Carraro},
  {Danziger}, {Fran{\c c}ois}, {Matteucci}, \& {Santos}}]{2009A&A...499L...9P}
{Pace}, G., {Melendez}, J., {Pasquini}, L., {et~al.} 2009, \aap, 499, L9

\bibitem[{{Perryman} {et~al.}(1997){Perryman}, {Lindegren}, {Kovalevsky},
  {Hoeg}, {Bastian}, {Bernacca}, {Cr{\'e}z{\'e}}, {Donati}, {Grenon},
  {Grewing}, {van Leeuwen}, {van der Marel}, {Mignard}, {Murray}, {Le Poole},
  {Schrijver}, {Turon}, {Arenou}, {Froeschl{\'e}}, \&
  {Petersen}}]{1997Perryman}
{Perryman}, M.~A.~C., {Lindegren}, L., {Kovalevsky}, J., {et~al.} 1997, \aap,
  323, L49

\bibitem[{{Queloz} {et~al.}(2001){Queloz}, {Henry}, {Sivan}, {Baliunas},
  {Beuzit}, {Donahue}, {Mayor}, {Naef}, {Perrier}, \& {Udry}}]{Queloz01}
{Queloz}, D., {Henry}, G.~W., {Sivan}, J.~P., {et~al.} 2001, \aap, 379, 279

\bibitem[{{Robertson} {et~al.}(2014){Robertson}, {Mahadevan}, {Endl}, \&
  {Roy}}]{2014Robertson}
{Robertson}, P., {Mahadevan}, S., {Endl}, M., \& {Roy}, A. 2014, Science, 345,
  440

\bibitem[{{Robinson} {et~al.}(1990){Robinson}, {Cram}, \&
  {Giampapa}}]{1990Robinson}
{Robinson}, R.~D., {Cram}, L.~E., \& {Giampapa}, M.~S. 1990, \apjs, 74, 891

\bibitem[{{Saar} {et~al.}(1998){Saar}, {Butler}, \&
  {Marcy}}]{1998ApJ...498L.153S}
{Saar}, S.~H., {Butler}, R.~P., \& {Marcy}, G.~W. 1998, \apjl, 498, L153

\bibitem[{{Saar} \& {Donahue}(1997)}]{1997ApJ...485..319S}
{Saar}, S.~H. \& {Donahue}, R.~A. 1997, \apj, 485, 319

\bibitem[{{Santos} {et~al.}(2010){Santos}, {Gomes da Silva}, {Lovis}, \&
  {Melo}}]{2010Santos}
{Santos}, N.~C., {Gomes da Silva}, J., {Lovis}, C., \& {Melo}, C. 2010, \aap,
  511, A54

\bibitem[{{Santos} {et~al.}(2000){Santos}, {Mayor}, {Naef}, {Pepe}, {Queloz},
  {Udry}, \& {Blecha}}]{2000A&A...361..265S}
{Santos}, N.~C., {Mayor}, M., {Naef}, D., {et~al.} 2000, \aap, 361, 265

\bibitem[{{Simpson} {et~al.}(2010){Simpson}, {Baliunas}, {Henry}, \&
  {Watson}}]{2010MNRAS.408.1666S}
{Simpson}, E.~K., {Baliunas}, S.~L., {Henry}, G.~W., \& {Watson}, C.~A. 2010,
  \mnras, 408, 1666

\bibitem[{{Skumanich}(1972)}]{1972ApJSkumanich}
{Skumanich}, A. 1972, \apj, 171, 565

\bibitem[{{Soderblom} {et~al.}(1993){Soderblom}, {Stauffer}, {Hudon}, \&
  {Jones}}]{1993ApJS...85..315S}
{Soderblom}, D.~R., {Stauffer}, J.~R., {Hudon}, J.~D., \& {Jones}, B.~F. 1993,
  \apjs, 85, 315

\bibitem[{{Strassmeier} {et~al.}(2000){Strassmeier}, {Washuettl}, {Granzer},
  {Scheck}, \& {Weber}}]{2000A&AS..142..275S}
{Strassmeier}, K., {Washuettl}, A., {Granzer}, T., {Scheck}, M., \& {Weber}, M.
  2000, \aaps, 142, 275

\bibitem[{{Strassmeier} {et~al.}(1990){Strassmeier}, {Fekel}, {Bopp},
  {Dempsey}, \& {Henry}}]{1990Stassmeier}
{Strassmeier}, K.~G., {Fekel}, F.~C., {Bopp}, B.~W., {Dempsey}, R.~C., \&
  {Henry}, G.~W. 1990, \apjs, 72, 191

\bibitem[{{Tody}(1993)}]{IRAF}
{Tody}, D. 1993, in Astronomical Society of the Pacific Conference Series,
  Vol.~52, Astronomical Data Analysis Software and Systems II, ed. R.~J.
  {Hanisch}, R.~J.~V. {Brissenden}, \& J.~{Barnes}, 173

\bibitem[{{Valenti} \& {Fischer}(2005)}]{2005ApJS..159..141V}
{Valenti}, J.~A. \& {Fischer}, D.~A. 2005, \apjs, 159, 141

\bibitem[{Vassallo(2014)}]{tesivassallo}
Vassallo, D. 2014, Master's thesis, Universita' degli Studi di Bologna, Italy

\bibitem[{{Vaughan} \& {Preston}(1980)}]{1980PASP...92..385V}
{Vaughan}, A.~H. \& {Preston}, G.~W. 1980, \pasp, 92, 385

\bibitem[{{Voges} {et~al.}(1999){Voges}, {Aschenbach}, {Boller},
  {Br{\"a}uninger}, {Briel}, {Burkert}, {Dennerl}, {Englhauser}, {Gruber},
  {Haberl}, {Hartner}, {Hasinger}, {K{\"u}rster}, {Pfeffermann}, {Pietsch},
  {Predehl}, {Rosso}, {Schmitt}, {Tr{\"u}mper}, \&
  {Zimmermann}}]{1999A&A...349..389V}
{Voges}, W., {Aschenbach}, B., {Boller}, T., {et~al.} 1999, \aap, 349, 389

\bibitem[{{Voges} {et~al.}(2000){Voges}, {Aschenbach}, {Boller}, {Brauninger},
  {Briel}, {Burkert}, {Dennerl}, {Englhauser}, {Gruber}, {Haberl}, {Hartner},
  {Hasinger}, {Pfeffermann}, {Pietsch}, {Predehl}, {Schmitt}, {Trumper}, \&
  {Zimmermann}}]{2000IAUC.7432R...1V}
{Voges}, W., {Aschenbach}, B., {Boller}, T., {et~al.} 2000, \iaucirc, 7432, 1

\bibitem[{{Wright} {et~al.}(2004){Wright}, {Marcy}, {Butler}, \&
  {Vogt}}]{2004ApJS..152..261W}
{Wright}, J.~T., {Marcy}, G.~W., {Butler}, R.~P., \& {Vogt}, S.~S. 2004, \apjs,
  152, 261

\bibitem[{{Zechmeister} \& {K{\"u}rster}(2009)}]{GLSP}
{Zechmeister}, M. \& {K{\"u}rster}, M. 2009, \aap, 496, 577

\bibitem[{{Zhao} {et~al.}(2011){Zhao}, {Oswalt}, {Rudkin}, {Zhao}, \&
  {Chen}}]{2011AJZhao}
{Zhao}, J.~K., {Oswalt}, T.~D., {Rudkin}, M., {Zhao}, G., \& {Chen}, Y.~Q.
  2011, \aj, 141, 107

\end{thebibliography}

\begin{appendix}
\section*{Stars with RV-\Ha\ correlation}
\label{app}
\begin{figure}[!h]
\includegraphics[width=\columnwidth]{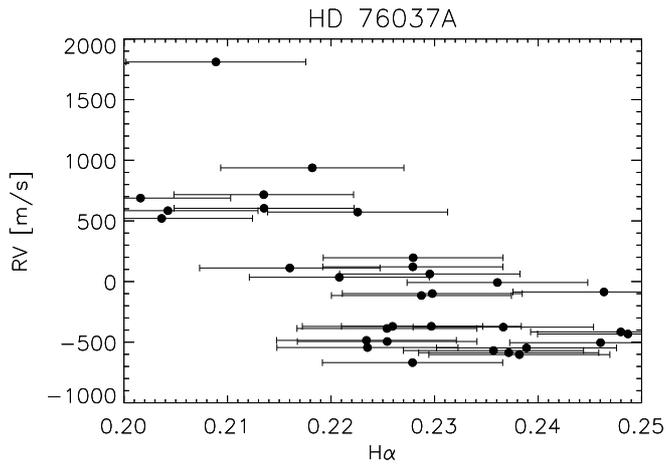}
\caption{RV as a function of the \Ha\ index for HD 76037A.}
\label{hd76037bHalphaGLS}
\end{figure}
\begin{figure}[!h]
\includegraphics[width=\columnwidth]{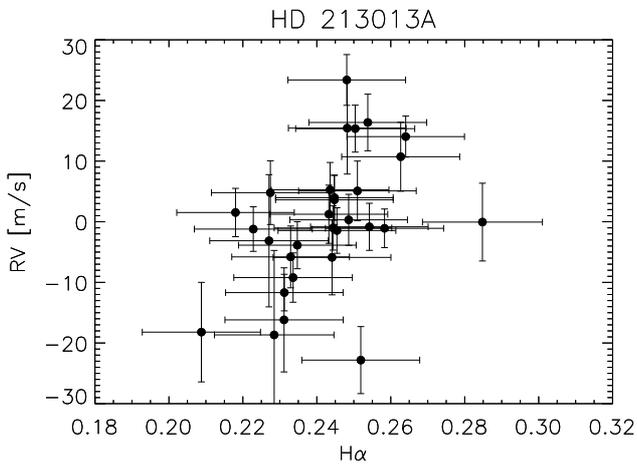}
\caption{RV as a function of the \Ha\ index for HD 213013A.}
\label{hd213013a_corr}
\end{figure}
\begin{figure}[!h]
\includegraphics[width=\columnwidth]{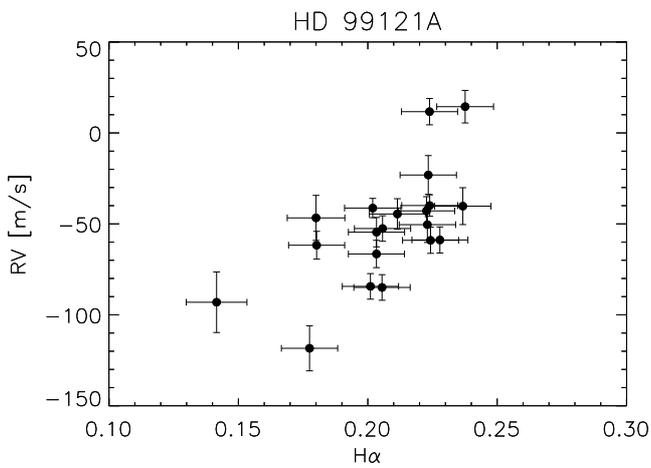}
\caption{Decontaminated RV as a function of the \Ha\ index for HD 99121A.}
\label{gj380_corr}
\end{figure}
\begin{figure}[!h]
\includegraphics[width=\columnwidth]{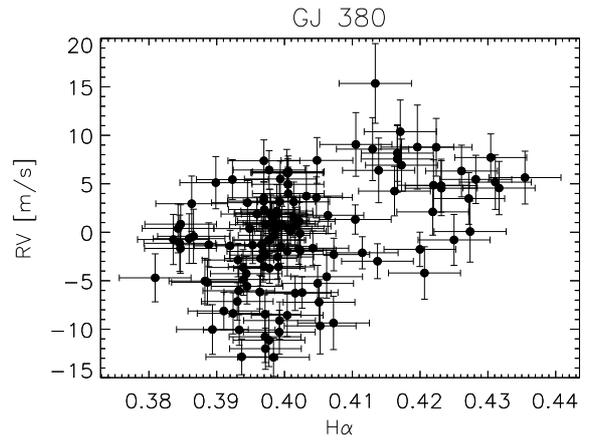}
\caption{RV as a function of the \Ha\ index for GJ 380.}
\label{gj380RV_corr}
\end{figure}

\end{appendix}

\end{document}